\documentclass[fleqn,usenatbib]{mnras}

\usepackage{newtxtext,newtxmath}
\usepackage[T1]{fontenc}

\DeclareRobustCommand{\VAN}[3]{#2}
\let\VANthebibliography\thebibliography
\def\thebibliography{\DeclareRobustCommand{\VAN}[3]{##3}\VANthebibliography}

\usepackage{graphicx}	% Including figure files
\usepackage{amsmath}	% Advanced maths commands
\usepackage{xfrac}
\usepackage{xspace}
\usepackage{caption}
\usepackage{subcaption}
\usepackage{soul}
\setstcolor{red}
\newcommand{\Abacus}{\textsc{Abacus}\xspace}
\newcommand{\AbacusSummit}{\textsc{AbacusSummit}\xspace}
\newcommand{\Rockstar}{\textsc{Rockstar}\xspace}
\newcommand{\FoF}{\textsc{fof}\xspace}
\newcommand{\CompaSO}{\textsc{CompaSO}\xspace}
\title[Convergence of halo statistics]{Convergence of halo statistics: code comparison between \Rockstar and \CompaSO using scale-free simulations}

\author[S. Maleubre et al.]{
Sara Maleubre,$^{1,2}$\thanks{E-mail: sara.maleubremolinero@physics.ox.ac.uk}
Daniel J.\ Eisenstein,$^{3}$
Lehman H.\ Garrison,$^{4,5}$
and Michael Joyce$^{1}$ \\
$^{1}$ Laboratoire de Physique Nucléaire et de Hautes Énergies, UPMC IN2P3 CNRS UMR 7585, \\ Sorbonne Université, 4, place Jussieu, 75252 Paris Cedex 05, France \\
$^{2}$ Sub-department of Astrophysics, University of Oxford, Keble Road, Oxford OX1 3RH, UK \\
$^{3}$ Center for Astrophysics $|$ Harvard $\&$ Smithsonian, 60 Garden St, Cambridge, MA 02138  \\
$^{4}$ Center for Computational Astrophysics, Flatiron Institute, 162 Fifth Ave., New York, NY 10010 \\
$^{5}$ Scientific Computing Core, Flatiron Institute, 162 Fifth Ave., New York, NY 10010 \\
}

\date{Accepted XXX. Received YYY; in original form ZZZ}

\pubyear{2022}

\begin{document}
\label{firstpage}
\pagerange{\pageref{firstpage}--\pageref{lastpage}}
\maketitle

\begin{abstract}
{
In this study, we perform a halo-finder code comparison between \Rockstar and \CompaSO. Based on our previous analysis aiming at quantifying resolution of \emph{N}-body simulations by exploiting large (up to $N=4096^3$) simulations of scale-free cosmologies run using \Abacus, we focus on convergence of the Halo Mass Function, 2-point Correlation Function and mean radial pairwise velocities of halo centres selected with the aforementioned two algorithms. 
%Here we focus on HMF, 2PCF and pairwise velocities of halo centres selected with two of the most popular halo-finders, \FoF and \Rockstar, together with the new \CompaSO algorithm. 
We establish convergence, for both \Rockstar and \CompaSO, of mass functions at the $1\%$ precision level and of the mean pairwise velocities (and also 2-point Correlation Function) at the $2\%$ level. At small scales and small masses, we find that \Rockstar exhibits greater self-similarity, and we also highlight the role played by the merger-tree post-processing of \CompaSO halos on their convergence. Finally, we give resolution limits expressed as a minimum particle number per halo in a form that can be directly extrapolated to LCDM.
}  
\end{abstract}


%%%%%%%%%%%%%%%%%%%%%%%%%%%%%%%%%%%%%%%%%%%%%%%%%%

%%%%%%%%%%%%%%%%% BODY OF PAPER %%%%%%%%%%%%%%%%%%

%----------------------------------------------------------------------
\section{Introduction}\label{sec:Intro}
Dark matter is thought to account for more than 85\% of the total matter in the universe. It forms clumps by gravitational attraction, channelling baryons together and serving as birthplaces for galaxies. Cosmological $N$-body simulations track dark matter particles, calculating their position and velocity at discrete time-steps. However, our central goal is to use them to produce predictions for observable objects, such as galaxies, and provide a framework for testing cosmological models. To this end, simulated dark matter particles are processed and bound together into large virialized objects, dark matter halos, whose evolution is expected to closely trace that of their hosted galaxies. Our procedure to identify such structures (halo-finding) requires certain choices on the definition of a halo: in particular, how its boundary is defined, where its centre is, or how particle membership is treated. As a consequence, the extracted properties of such objects are very sensitive to the particular model used to define these structures, and the precision of most statistical measurements is much poorer than for the dark matter field.

The issue of the precision of relevant halo properties is particularly complex, as it combines two distinct issues: that of the precision with which mass distribution in the $N$-body simulation represents the physical limit \cite[previously analysed in][]{Joyce2021,Maleubre2022,Maleubre2023}, and that of the halo definition and extraction (which will be the focus of this paper). Numerous ways have been proposed and exploited \cite[see][for a review]{Knebe2013}, but still different halo-finders running on the same simulation indeed show different results \citep{Knebe2011}. This illustrates the previous statement, indicating that a large fraction of the uncertainty in retrieving information about halo properties is actually due to the process of halo finding itself. We will study here the accuracy at which we can measure different halo properties and statistics when using two different halo finders (\Rockstar and \CompaSO).

Here, we use the techniques introduced in \citet{Joyce2021} and developed and 
applied also in \citet{Leroy2021,Garrison2021,Garrison2021c} and \citet{Maleubre2022} to derive resolution limits  
arising from particle discretization for different halo statistics by analysing deviations from self-similarity in scale-free cosmological models. In particular, we expand the analysis in \citet{Maleubre2023} of the radial component of the pairwise velocity in the matter field to halos, as well as to assess and quantify the limits on the precision with respect to halo-finder. In addition, we revisit and develop further the analysis in \cite{Leroy2021} of the 
mass functions and two-point correlation function of halos (that uses \FoF and \Rockstar halo-finders), extending the comparison  
to now include the new halo finder \CompaSO \citep{CompaSO,Sownak2022}, as well as both larger simulations and scale-free models with different exponents.
We underline that our methods allow us to test each halo finder individually for their convergence properties, but does not allow us to conclude whether one is better than the other for constructing observables (as this largely depends on the observable and the scales of interest), or which halo definition is the more physically relevant one. What we can establish are the limits in mass and scale at which halos found using a particular halo finder, and some of their relevant statistics, aren't affected by unphysical scales at a certain precision.
% We point out that we only test each halo-finder individually for their convergence properties, but we cannot say if one is better than the other for constructing observables, or which halo definition is the "correct one". 

This article is structured as follows. The first part of \autoref{sec:PWandSF} recaps what scale-free cosmologies are and how their self-similar evolution can be used to determine the accuracy at which different statistics can be measured in $N$-body simulations. We end the section with a description of the halo statistics that will be analysed. \autoref{sec:NumSim} contains a summary of the simulations used, as well as a brief description of \Abacus, the $N$-body code used for their computation. It also summarizes the methods used to estimate convergence of the different statistics, and ends with a summary of the halo finders we compare (\Rockstar and \CompaSO). In \autoref{sec:Results} we present and analyse our results, and finally, we summarize them in \autoref{sec:Conclusions}.

%----------------------------------------------------------------------
\section{Scale-Free Simulations and Halo Statistics}\label{sec:PWandSF}
\subsection{Scale-free simulations and Self-Similarity}

The self-similarity of scale-free models has been widely exploited since the early development of $N$-body simulations, as an instrument to check the reliability of results \citep[e.g][]{efstathiou1988gravitational, colombi_etal_1996}, and study halo properties \citep[e.g][]{Cole1996, Navarro1997, Knollman2008, Elahi2009, Diemer2015, Ludlow2016, Diemer2019}. Following our previous investigations \citep{Joyce2021, Leroy2021, Garrison2021,Maleubre2022,Maleubre2023} we will analyse self-similarity of scale-free cosmologies to extract quantitative constraints on resolution for different halo-finders.

In scale-free simulations, the initial power spectrum of fluctuations is a power law of the form $P(k)\propto k^n$, where the spectral index \emph{n} is fixed for each cosmology. They have an Einstein de Sitter (EdS) background ($\Omega_{\text{tot}}=\Omega_M=1$) following an expansion law $a(t)\propto t^{2/3}$, and thus are characterized by just one scale, the scale of non-linearity. This length scale may be defined by
\begin{equation}\label{eq:RNL}
    \sigma_{\text{lin}}^2(R_{\text{NL}},a)=1
\end{equation}
where $\sigma^2_{\rm lin}$ is the variance of normalized linear mass fluctuations in a sphere at a given time. Using linear perturbation theory we infer
\begin{equation}\label{eq:RNL-scaling}
    R_{\text{NL}}\propto a^{\frac{2}{3+n}}
\end{equation}
which gives the relation between time and scale in these types of cosmological models.

From \autoref{eq:RNL-scaling} we can deduce that, in the absence of additional independent length scales, clustering evolution must behave self-similarly. %i.e. a single solution describes the structure and time behaviour of the system, when expressed in properly scaled variables. 
This means that for any dimensionless clustering statistic, its scale and time dependence can be simplified to
\begin{equation}\label{eq:SS-sgeneral}
    F(x_1,x_2,...;a) = F_0(x_i/X_{\text{NL},i}(a))
\end{equation}
where each $X_{\text{NL},i}(a)$ encodes the temporal dependence of any quantity with the dimensions of $x_i$, inferred from self-similar rescaling.

From \autoref{eq:RNL} and \ref{eq:RNL-scaling} we can define the rescaling quantities used in the current analysis, the characteristic length and mass scales of non linearity. Defining $\sigma_i\equiv\sigma_{\text{lin}}(\Lambda,a_i)$ where $a_i$ is the value of the scale factor at the start of the simulation, we can infer 
\begin{equation}
    R_{\text{NL}}(a)=\Lambda\left(\frac{a}{a_i}\sigma_i\right)^{2/(3+n)}
\end{equation}
and subsequently
\begin{equation}\label{eq:MNL}
    M_{\text{NL}}(a)=\frac{4\pi}{3}\bar{\rho}R_{\text{NL}}^3(a)=\frac{4\pi}{3}m_{P}\left(\frac{a}{a_i}\sigma_i\right)^{6/(3+n)}
\end{equation}
where $\bar{\rho}$ is the mean (comoving) mass density and $m_P=\Lambda^3\bar{\rho}$ is the mass of a particle in the simulation (with $\Lambda=L/N^{1/3}$ the mean inter-particle spacing of the initial grid).

In our analysis, as we have been doing for the previous studies, we will use this property of self-similarity to assess the range of scales that a simulation can reproduce at a desired precision, for some given statistic. In particular, this work will treat the reliability of different halo-finders, notably how the resolved scales depend on a halo's particle number.

\subsection{Halo quantities}\label{sec:Halo_quant}

In \autoref{subsec:R_CSO_stats} we will use self-similarity to test two different halo selection algorithms (\Rockstar and \CompaSO).

We will start by analysing the convergence of the mass function (HMF) as a function of rescaled mass, as clustering statistics are measured as a function of the mass of halos. We recall that the HMF is just the number density of halos of a given mass at a given redshift. Following the treatment of Press $\&$ Schechter \citep{PressSchechter1974}, it is convenient to express it in terms of the ``multiplicity'' function $f$ \citep{Jenkins2000,Tinker2008} defined by
\begin{equation}
    \frac{dn_{\text{h}}}{d\ln M}=f(\sigma_{\rm lin})\frac{\bar{\rho}}{M}\frac{d\ln\sigma_{\rm lin}^{-1}}{d\ln M}
\end{equation}
where $\bar{\rho}$ is the mean matter density, and $\sigma_{\rm lin}^2$ is expressed as a function of mass using $M_{\rm NL}\propto R_{\rm NL}^{3}$. 

For scale-free cosmologies, we can conveniently write $f(\sigma_{\rm lin})$ in terms of the rescaled mass $M/M_{\rm NL}$ as
\begin{equation}\label{eq:f}
    f(M/M_{\rm NL}) = \frac{6}{3+n}\frac{M^2\tilde{n}_{\text{h}}}{\bar{\rho}}
\end{equation}
where we've defined $\tilde{n}_{\rm h}\equiv dn_{\text{h}}/dM$ and from \autoref{eq:RNL-scaling} $d\ln\sigma^{-1}_{\rm lin}/d\ln M=(3+n)/6$. We refer hereafter to $f$ as the \emph{halo mass function} (HMF). 

We will continue our analysis with the halo-halo 2PCF, $\xi_{\rm hh}(r,M,a)$, which is a dimensionless function of the separation $r$ and of the halo mass $M$, calculated at a given snapshot:
\begin{equation}
   \xi_{hh}(r,M,a)= \left. \left<\delta n_h(0) \delta n_h(r) \right> \right|_{M,a}
\end{equation}
   where $\delta n_h$ is the fluctuation in the halo number density.

Similarly, the radial component of the pairwise velocity can be computed as the correlation between two centres weighted by their projected velocity
\begin{equation}
    v^{r}_{12}=\left<(\mathbf{v}_1-\mathbf{v}_2)\cdot\frac{\mathbf{r}}{|\mathbf{r}|}\right>
\end{equation}
where the velocity difference $(\bf{v}_1-\bf{v}_2)$ of a pair of halo centres is projected on to their separation vector $\bf{r}$, and $<\cdots >$ denotes the ensemble average.

In both cases, if self-similarity applies, these statistics are conveniently rewritten 
in terms of the dimensionless rescaled functions as
\begin{equation}
    \xi_{\rm hh}(r,M,a) = \xi_{{\rm hh}}(r/R_{\rm NL},M/M_{\rm NL})
\end{equation}
\begin{equation}
    \frac{v_{r,{\rm hh}}}{Hr} = V_{r,hh}(r/R_{\rm NL},M/M_{\rm NL}) \,.
\end{equation}

Following the procedure introduced in \citet{Maleubre2023}, we have used a modification of the analysis tool \emph{Corrfunc} \citep{Corrfunc_extra,Corrfunc2020} to calculate both the 2PCF and the radial component of the pairwise velocity.

%----------------------------------------------------------------------
\section{Numerical simulations}\label{sec:NumSim}
\subsection{{\mdseries\Abacus} code and simulation parameters}
We report results based on the simulations listed in \autoref{tab:example_table}. We make use of the \Abacus $N$-body code \citep{Garrison2021b}, which offers high performance and accuracy. It is based on CPU calculations of the far-field forces by a high-order multipole expansion, and an accelerated GPU calculation of near-field forces by pairwise evaluation. The $N=1024^3$ simulations were run using local facilities at the Harvard-Smithsonian Center for Astrophysics (CfA), while the larger $N=4096^3$ simulations are part of the \textsc{AbacusSummit} project \citep{Maksimova2021}, which used the Summit supercomputer of the Oak Ridge Leadership Computing Facility.

In this work we analyse two different exponents ($n=-1.5$, $n=-2.0$), relevant to standard (i.e. LCDM-like) models.
We use simulations of two different sizes ($N$, $L$) but otherwise identical parameters, allowing us to study finite box size effects. For the larger ($N=4096^3$) simulations, the statistics have been calculated on (random) sub-samples of different sizes (25\%, 3\%) to facilitate the assessment of finite sampling effects.

We work in units of the mean inter-particle (i.e. initial grid) spacing, $\Lambda=L/N^{1/3}$, and of the particle mass of the simulation, $m_P=\Lambda^3\bar{\rho}$. The essential time-stepping parameter in
\Abacus has been chosen as $\eta=0.15$  for all simulations, and 
the additional numerical parameters have been set as detailed in \citet{Maleubre2022} and summarized below. These choices are based on the extensive convergence tests of these parameters reported in our previous studies \cite[see also][]{Joyce2021,Garrison2021}. 

As in our previous studies, the Gaussian initial conditions are specified at a time $a=a_i$ fixed by the value of top-hat fluctuations at the particle spacing
\begin{equation}
    \sigma_{\rm lin}(\Lambda, a_i)=0.03
\end{equation}
They are set up using a modification to the standard Zel'dovich approximation as detailed in \citet{Garrison2016}. They apply a correction described by particle linear theory (PLT) as reported in \citet{Joyce2007}, as well as second order Lagrangian perturbation theory (2LPT) corrections. The former corrects the initial conditions for discreteness effects at early times, so that linear theory evolution is exact at a target time $a=a_{\text{PLT}}$. For all our simulations, this time has been chosen to coincide with the first output epoch as described below.

This first output epoch ($a=a_0$), corresponds to 
\begin{equation}
    \sigma_{\rm lin}(\Lambda,a_0)=0.56
\end{equation}
i.e. approximately the time of formation of the first non-linear structures, when fluctuations of peak-height $\nu\sim 3$ are expected to virialize in the spherical collapse model ($\sigma\sim\delta_c/\nu$, with $\delta_c=1.68$). This time has also been chosen to coincide with the target time of the PLT corrections $a_{\text{PLT}}$.

Subsequent output values are spaced by a factor $\sqrt{2}$ in the non-linear mass scale. Plugging this into \autoref{eq:MNL}, we get:
\begin{equation}\label{eq:diff_a}
    \Delta\log_2 a = \frac{3+n}{6}\Delta\log_2 M_{\rm NL}=\frac{3+n}{12} 
\end{equation}

In practice, we use $\log_2(a/a_0)$ as the time variable of our analysis, which indicates how many epochs have passed since the first output, but internally we also make used of a variable $S=0,1,2,...$, corresponding to the different outputs of the simulation, with
\begin{equation}\label{eq:Sdef}
    S = \frac{12}{3+n}\log_2\left(\frac{a_S}{a_0}\right)
\end{equation}

The force softening of all simulations has been fixed in physical coordinates (evolving as $\epsilon(a)\propto 1/a$ in comoving ones), and taking the value $\epsilon(a_0)/\Lambda=0.3$. Nevertheless, a fixed comoving softening $\epsilon(a<a_0)=0.3$ was also imposed to avoid excessively large softening values at earlier times, down to the first output of the simulation. This has been previously tested in \citet{Garrison2021} and \citet{Maleubre2022}, and shown to provide both accuracy and computational efficiency for the present spectral indices.

\begin{table*}
	\centering
	\caption{Summary of the $N$-body simulation data used for the analysis of this paper. The first column shows the spectral index of the initial PS, and $N$ is the number of particles of each simulation. The third column gives the ratio of the effective Plummer force smoothing length $\epsilon$ to mean inter-particle separation (equal to the initial grid spacing $\Lambda$), at the time of our first output. This smoothing is fixed in proper coordinates. The last column indicates the halo-finder utilized.}
	\label{tab:example_table}
    \begin{tabular}{cccccc}
    \hline
    $n$       & $N$       & $\epsilon_0/\Lambda$    & Halo Finder     \\ \hline \hline
    $n=-1.5$  & $4096^3$  & 0.3                     & \CompaSO      \\
    $n=-1.5$  & $1024^3$  & 0.3                     & \Rockstar     \\ \hline
    $n=-2.0$  & $4096^3$  & 0.3                     & \CompaSO      \\
    $n=-2.0$  & $1024^3$  & 0.3                     & \Rockstar     \\ \hline
    % $n=-2.25$ & $4096^3$ & 0.3                & 1          & 3\%            & $\xi$ /                               \\
    \end{tabular}
\end{table*}

\subsection{Estimation of converged values}\label{sec:converge_method}

As in our previous papers, the convergence to the physical limit of the targeted statistics will be studied by analysing the behaviour of their temporal evolution, which becomes time-independent, in rescaled coordinates, in the case of self-similarity. Our final objective is to perform a quantitative analysis of convergence --- i.e. to identify estimated converged values, and converged regions at some precision --- for which we need to fix a criterion. The conclusions drawn should not depend significantly on the method, as can be inferred by comparing the results for the simulation with $n=-2.0$ and $N=1024^3$ analysed with \Rockstar in this paper with those of an earlier study in \citet{Leroy2021}, which used slightly different criteria~\footnote{\citet{Leroy2021} used a simulation with a different force softening, shown to be converged, for matter statistics, over a lesser range of scales by \citet{Garrison2021} and \citet{Maleubre2022} than the one used in this study. Subsequently, small convergence disparities can be attributed to the simulation rather than the method used to estimate accuracy.}. The described methods will be equivalent for all halo statistics ($f(M/M_{\rm NL})$, $\xi_{\rm hh}$, $v_{r,{\rm hh}}/Hr$), so we will denote them by $X$ in the following.

The criterion imposed for the current analysis follows that presented in \citet{Maleubre2022}, and further used in \citet{Maleubre2023}, which we will recap here. It will allow us to estimate both a converged value and a converged temporal region at a chosen precision, per rescaled bin in mass (and halo-separation) for each of the relevant statistics.

We start by calculating an estimated converged value, $X_{\text{est}}$, as the average of the statistic in a fixed-size temporal window minimizing
\begin{equation}
    \Delta = \frac{\left|X_{\text{max}}-X_{\text{min}}\right|}{2\mu_{X}}
\end{equation}
where $X_{\text{max}}$, $X_{\text{min}}$ and $\mu_{X}$ are, respectively, the maximum, minimum, and average values in the window. We say that a bin is converged at precision $p$ if the minimum value of $\Delta$ is less than $p$.

We note that $X_{\text{est}}$ is calculated in a fixed temporal window, and it is only used to assess whether the statistic is converged it a particular value of the rescaled variable. For the purpose of this study, the width of this temporal window was chosen to be $w=5$, where $w$ represent the number of consecutive snapshots, such that the non-linearity scale increases by $M_{\text{NL}}\sim 2^{w/2}$ ($R_{\text{NL}}\sim 2^{w/6}$). The choice for the size of this window is somewhat arbitrary, but the results should not significantly depend on it. A window that is chosen to be too small will lead to "false positives" in convergence, and contradictory results with the step 2 explained below. On the other hand, a window which is too large will exclude smaller but apparently converged temporal regions.

In a second step, we define the entire region of convergence by finding the largest connected temporal window (containing at least three consecutive snapshots, though again this number is not crucial) verifying
\begin{equation}\label{eq:conv_criterion}
    \frac{\left|X-X_{\text{est}}\right|}{X_{\text{est}}}<p
\end{equation}
This allows us to correct for missed converged snapshots in step one, by finding all points "close" (within the required region of accuracy) to the estimated converged value $X_{\text{est}}$.

The reported converged value of the statistic at each rescaled bin is then calculated as the mean value of the statistic within the full resolved region (i.e. the region verifying the condition in \autoref{eq:conv_criterion}), and it is denoted by $X_{\text{conv}}$. In this case, the precision at which the statistic is evaluated in the simulation is given by $p$, and when we say that we have a precision at $x\%$ we mean that $p=x/100$.

While choosing the binning of the different studied statistics, we ensure that they match at different snapshots when rescaled by $M_{\text{NL}}$ ($R_{\text{NL}}$), to facilitate the comparison between them. We use constant logarithmic spacing such that $1+(\Delta m/m)\approx 2^{1/3}$ and $1+(\Delta r/r)\approx 2^{1/12}$. In order to reduce statistical noise sufficiently, we have rebinned by grouping two (four) such bins, corresponding to $\Delta m/m\approx 0.55$ ($\Delta r/r \approx 0.26$). All bins in the rescaled variables reported in the results below are labelled by their geometrical centre.

\subsection{Halo Finders: {\mdseries\Rockstar} and {\mdseries\CompaSO}}\label{sec:HaloFinder}

In this paper we analyse results from two different group-finding algorithms, comparing their resolution for a set of halo-statistics, as well as the accuracy of convergence. COMPetitive Assignment to Spherical Overdensities \cite[\CompaSO,][]{CompaSO} is a newly developed halo-finder specifically created to meet the demanding requirements of the \AbacusSummit cosmological $N$-body simulations. It runs on-the-fly, as part of the simulation code itself, with two of its primary requirements being keeping up with the high speed of \Abacus \citep{Maksimova2021}, and supporting the creation of catalogues and merger trees to be used in the Dark Energy Spectroscopic Instrument \citep[DESI,][]{DESI} project. On the other hand, Robust Overdensity Calculation using K-Space Topologically Adaptive Refinement \cite[\Rockstar,][]{Rockstar} is a well established, widely used halo-finding algorithm, which uses information from both position and velocity of the particles.

The \CompaSO algorithm is a configuration-space, FoF and SO algorithm to compute halos from $N$-body simulations. It first obtains a measurement of the local density using a kernel of the form $W = 1-r^2/b^2_{\text{kernel}}$, where typically $b_{\text{kernel}}=0.4\Lambda$. Particles are linked together into FoF groups (L0 halos) as long as their local density ($\Delta$) is higher than a chosen threshold. The main halos (L1 halos) are then formed inside these groups. Within each group, the algorithm finds the particle with the highest kernel density---the first halo nucleus---and makes a preliminary assignment to it of all particles within a radius $R_{\rm L1}$ (innermost radius enclosing $\Delta<\Delta_{\rm L1}=200$ in EdS). Particles outside $80\%$ of $R_{\rm L1}$ are eligible to become their own halo centre as long as they are the densest within their kernel radius. The algorithm then finds the next highest density among eligible particles, which becomes the next halo nucleus. Particles are assigned to the first nucleus, but if a particle belongs to two halos, the algorithm performs a competitive assignment. This reassigns a particle to a new halo if its enclosed density with respect to the new halo is at least twice that of the old one. The search for new halo centres within L0 continues until no particles remain that are likely to nucleate halos of sufficient density.

\CompaSO can sometimes fragment elongated halos into multiple objects, due to its spherical nature, or identify substructure as a distinct halo at one epoch that was already identified as a monolithic halo at a previous epoch.  For this reason, a cleaning procedure is performed in post-processing, relying on merger-trees information \cite{Sownak2022}. This procedure checks what fraction of the particles of a halo at time $t_i$ come from a much larger halo located at a similar position at time $t_{\rm i-1}$ and $t_{\rm i-2}$. If a sufficiently large fraction did, then the newer halo is deemed a “potential split” and merged into the larger halo. In addition, if at an earlier redshift a halo peak mass exceeds more than twice its present day mass, it is also merged into a more massive neighbour, from whom it had presumably split off. 

We will be particularly testing how this procedure improves halos determined by \CompaSO. The described cleaning method affects, in general, low-mass halos around more massive ones, appending their particle list to the latter, and resulting in cleaned halo catalogues with a lower number of smaller halos vs. a larger number of bigger halos \citep[][reported that this post-processing method removes ~1-5\% of objects]{Sownak2022}. As we will show in \autoref{subsec:R_CSO_stats}, this shifts the value of the HMF in each mass-bin exactly in the correct direction to preserve self-similarity, which is evidence for the good performance of the procedure.

\Rockstar is a group-finding method that uses information from both the position and the velocity of the particles. It works in a six dimensional phase-space framework, with an optional time refinement algorithm that tracks mergers. The code starts by creating FoF groups of a linking length larger than standard ($b=0.28$ by default), which assures that virial spherical overdensities can be determined within. For each of these FoF groups, a phase-space metric is defined by normalizing the positions and velocities of the particles by the position ($\sigma_x^2$) and velocity ($\sigma_v^2$) dispersions of the group, such that for two particles $p_1$ and $p_2$ the distance metric is defined by:
\begin{equation} \label{eq:metricR}
    d(p_1,p_2)=\left(\frac{|\mathbf{x}_1-\mathbf{x}_2|^2}{\sigma_x^2}+\frac{|\mathbf{v}_1-\mathbf{v}_2|^2}{\sigma_v^2}\right)^{1/2}
\end{equation}
The algorithm now performs a modified FoF in phase-space within each group, where it links particles with and adaptive phase-space linking length such that a constant fraction of them (default $70\%$) is always linked together with at least another particle into subgroups. The process repeats for each subgroup, creating a hierarchical set of structures until a minimum size substructure is found at the deepest level. Seed halos are placed at this final structure, and particles at higher levels are assigned to the closest seed halo in phase-space, where now the metric (\autoref{eq:metricR}) is calculated with respect to the seed halo. More than one seed can be found within each of the first level FoF groups, corresponding to either a halo or subhalo. This categorization is performed by including temporal information of previous steps, following particle-halo associations across time-steps. During its final step, \Rockstar calculates the gravitational potential of all particles using a modified Barnes-Hut method in order to unbind particles. 

\Rockstar defines halo masses by using various (user-specified) Spherical Overdensity (SO) criteria. We generally (except when stated otherwise) restrict ourselves in this study to using the SO mass corresponding to the virial radius, including all halo structures and considering only gravitationally bound mass \texttt{STRICT\_SO\_MASSES=0}\footnote{See \citep{Leroy2021} for a study exploring the differences to the HMF analysis arising from using bound vs. unbound particles, as well as parents only, subhalos, and all structures.}. All halo centres and velocities are calculated using a subset of the innermost particles ($\sim10\%$ of the halo radius), minimizing a Poisson error $\sigma/\sqrt{N}$. Finally, the algorithm has been run using \emph{default} parameters but for: \\ \texttt{TEMPORAL\_HALO\_FINDING = 0} and \texttt{MIN\_HALO\_OUTPUT\_SIZE = 25}.

%----------------------------------------------------------------------
\section{Results}\label{sec:Results}
We will start with the analysis of self-similarity of the HMF, and then explore the convergence of the radial pairwise velocities of halos, comparing it also with that of the 2PCFs. The analysis will compare results from \Rockstar and \CompaSO halos, with the latter focused on "cleaned" vs. "un-cleaned" results, for the different simulations reported in \autoref{tab:example_table}.  

\subsection{Halo mass function}\label{subsec:R_CSO_stats}

\begin{figure*}
    \begin{subfigure}{\linewidth}%{\textwidth}
    % \centering
    \includegraphics[width=\linewidth]{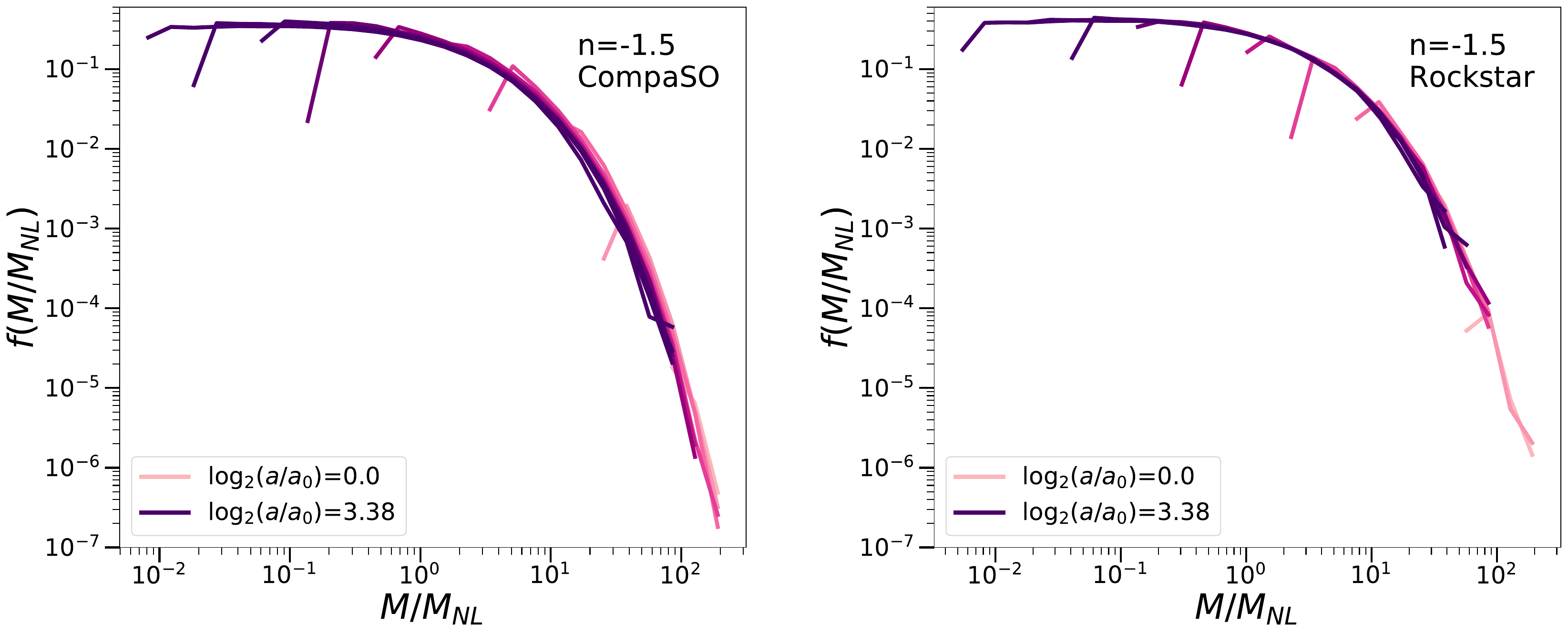}
    \end{subfigure}
    \begin{subfigure}{\linewidth}%{\textwidth}
    % \centering
    \includegraphics[width=\linewidth]{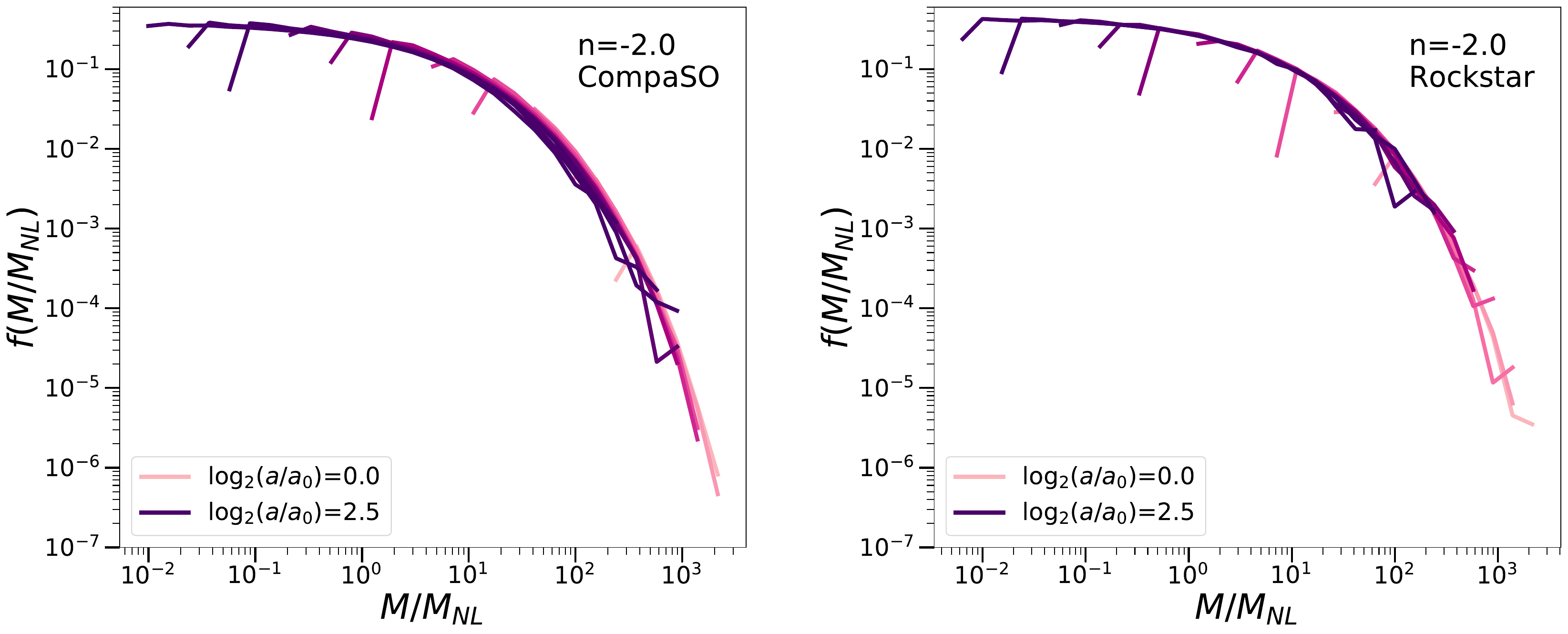}
    \end{subfigure}    
\caption{HMF as a function of rescaled mass $M/M_{\text{NL}}$ for simulations of spectral index $n=-1.5$ (upper panels) and $n=-2.0$ (lower panels). The left column shows results from \CompaSO \protect\cite[after the cleaning procedure detailed in][]{Sownak2022} for the $N=4096^3$ simulation. The right column shows results from \Rockstar for a single $N=1024^3$ simulation. Self-similar evolution corresponds to the superposition of the curves in each plot. The times shown correspond to every third snapshot $S=0,3,6,... $ over the total span of the simulation.}
\label{fig:HMF2}
\end{figure*}

%We refer hereafter to $f$ as the \emph{halo mass function} (HMF). 
We first show in \autoref{fig:HMF2} the HMF as a function of rescaled mass $M/M_{\text{NL}}$, as defined in \autoref{sec:Halo_quant}. The two left panels correspond to the \CompaSO catalogue obtained from the $N=4096^3$ simulations of $n=-1.5$ (upper) and $n=-2.0$ (lower), while the right panels are for the \Rockstar catalogues of 
a $N=1024^3$ simulation with the same two exponents. All plots show that the self-similar rescaling appears to apply to a good approximation, especially at late times. The smaller \Rockstar boxes show greater deviations at larger rescaled masses, which are simply due to the reduced number of halos in the smaller volume. Comparing the two indices for each halo finder, we observe that the self similarity for $n=-2.0$ suffers small deviations (more notably in the smaller boxes analysed with \Rockstar) at smaller scales and times than $n=-1.5$ (look at halos beyond $10^3$ particles). This difference mirrors what we observed in previous analysis for dark matter statistics \citep{Joyce2021,Maleubre2022,Maleubre2023}, and reflects the increasing importance of finite box size as the index of the initial power spectrum reddens.

We next study these qualitatively-apparent deviations from convergence quantitatively by considering 
vertical slices in \ref{fig:HMF2}, assessing the self-similarity of the HMF as a function of time in bins of fixed rescaled mass. 

%Results for the comparison between \FoF and \Rockstar are shown in \autoref{fig:FoF} and \ref{fig:Rockstar}, respectively. Each panel shows the measured value of the rescaled mass function at all available snapshots of the simulation, and for a finite bin of halo mass $M/M_{\text{NL}}$. These bins have an equal logarithmic spacing with a width of about $40\%$ their central value and are equally spaced across the full range sampled by the simulation. To facilitate comparison between the two figures/halo-finders, their $y$-axis is fixed such that $y_{\text{max}}/y_{\text{min}}=2$. The dashed vertical lines in each plot indicate a particle resolution corresponding to 50 and 5000 particles, for the geometric centre of the mass-bin. In addition, \autoref{fig:Rockstar} has a small plot showing the fractional change $\Delta Y/Y$ (with $Y=M^2_{\text{NL}}n(M,a))$ between consecutive snapshots.

\begin{figure*}
    \begin{subfigure}{0.49\linewidth}%{0.24\textwidth}
    \includegraphics[width=\linewidth]{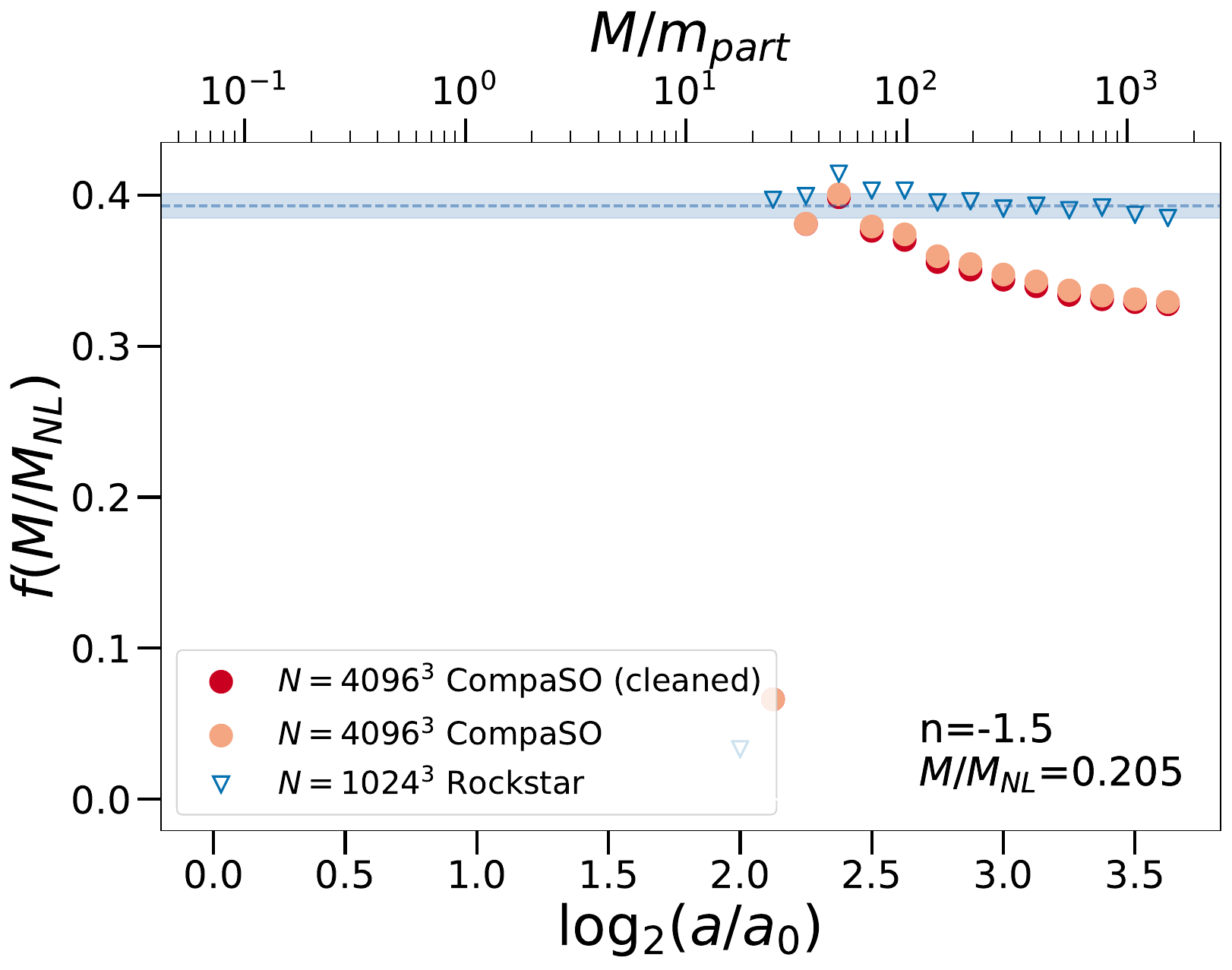}
    \end{subfigure}
    \begin{subfigure}{0.49\linewidth}%{0.24\textwidth}
    \includegraphics[width=\linewidth]{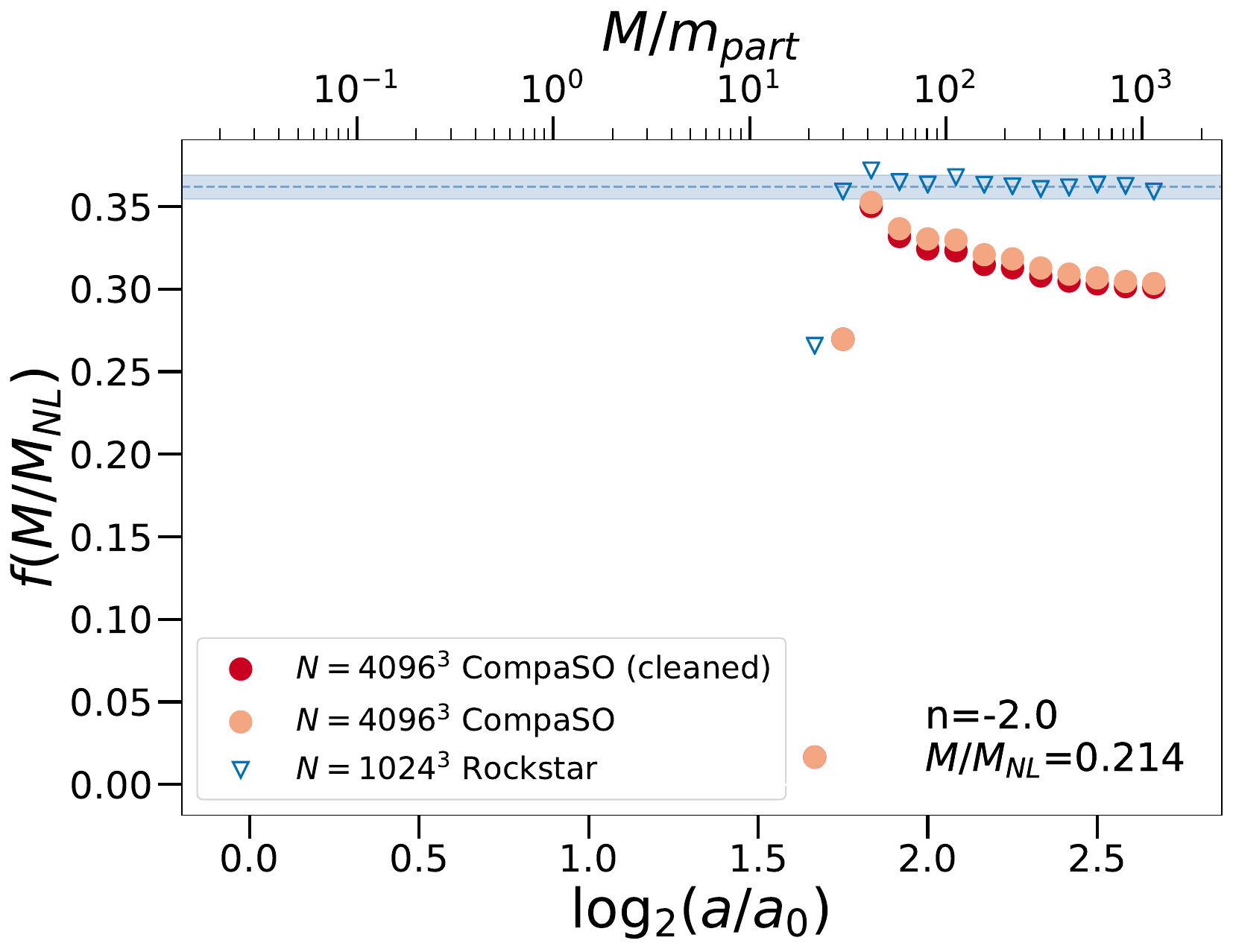}
    \end{subfigure}    
    \begin{subfigure}{0.49\linewidth}%{0.24\textwidth}
    \includegraphics[width=\linewidth]{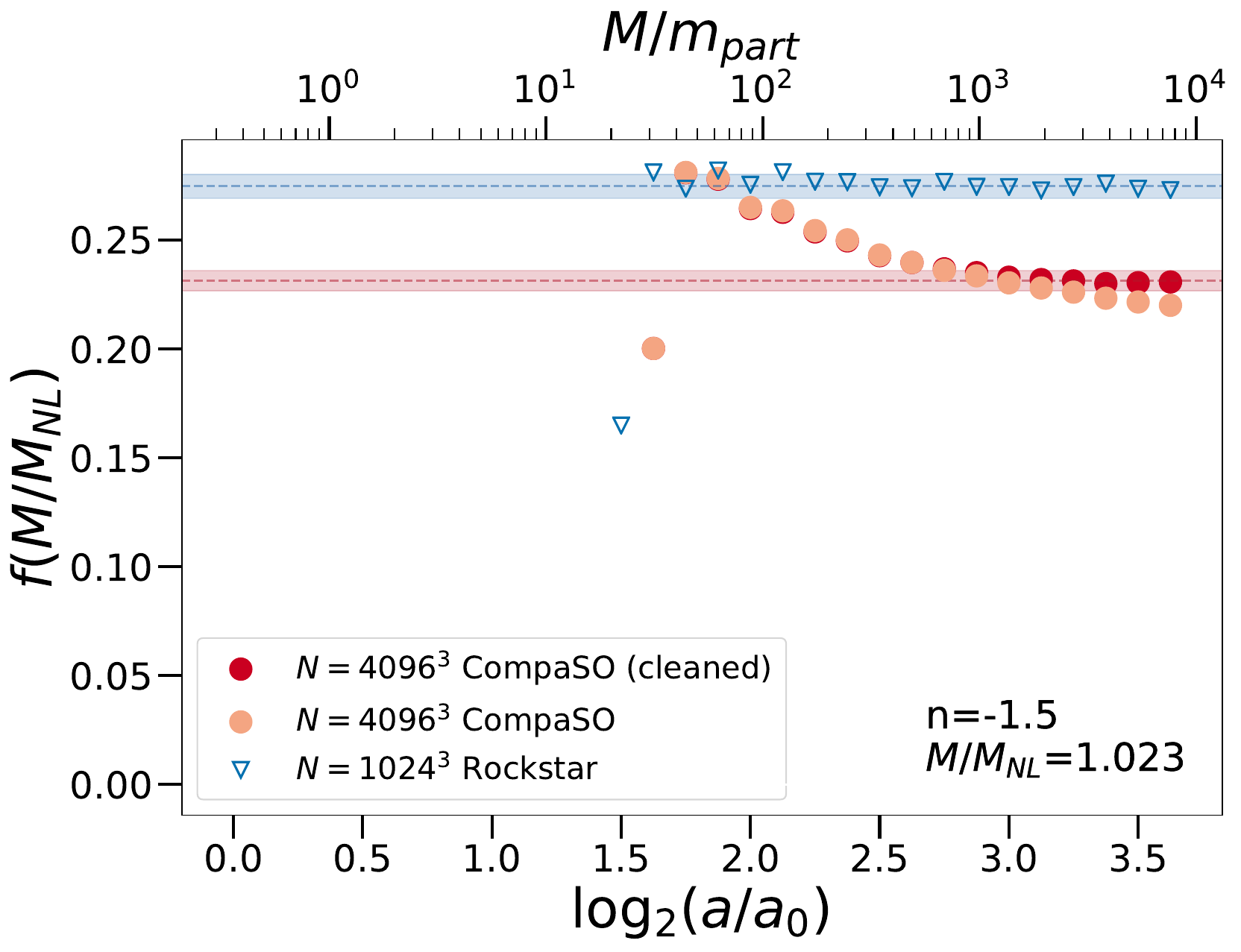}
    \end{subfigure}    
    \begin{subfigure}{0.49\linewidth}%{0.24\textwidth}
    \includegraphics[width=\linewidth]{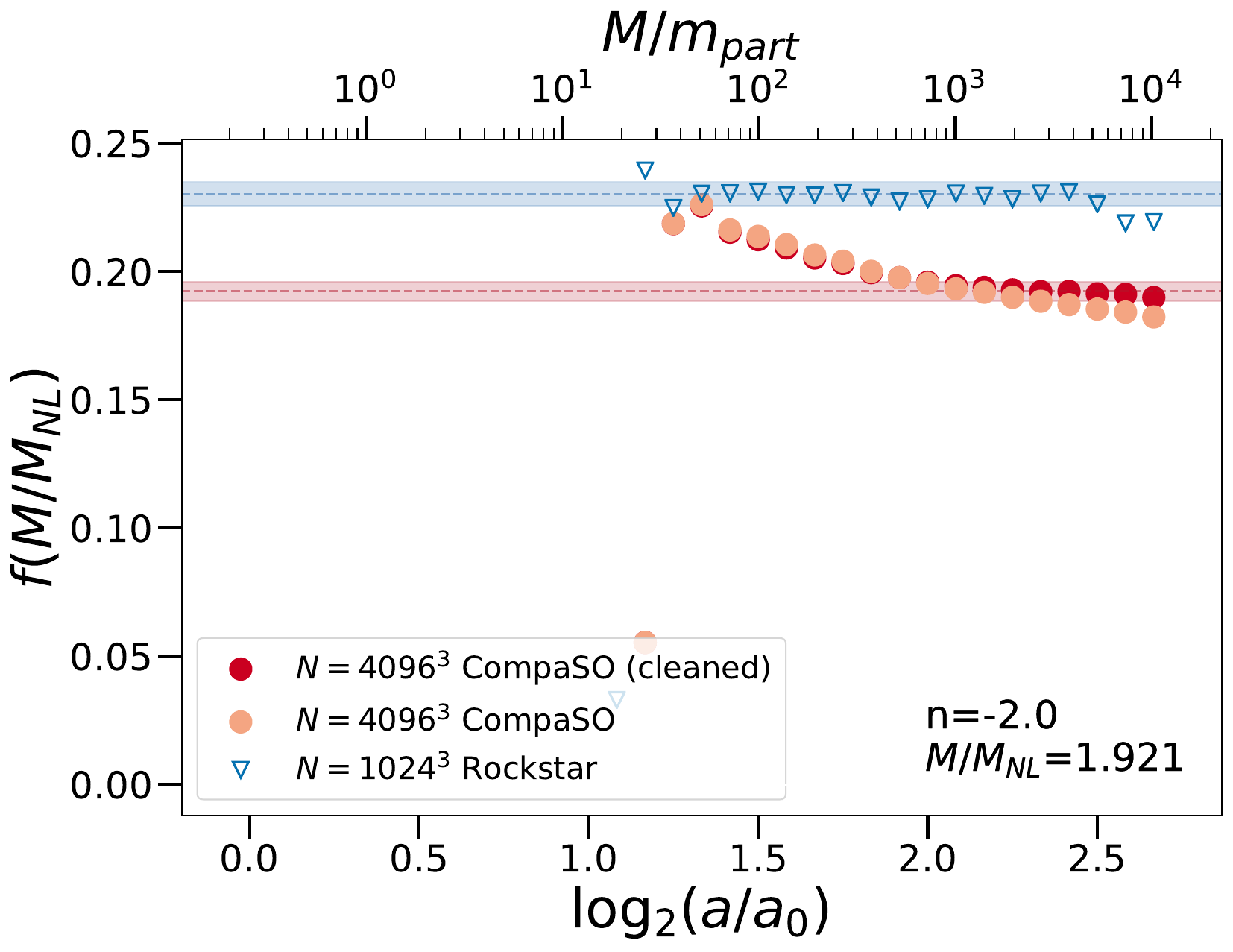}
    \end{subfigure}    
    \begin{subfigure}{0.49\linewidth}%{0.24\textwidth}
    \includegraphics[width=\linewidth]{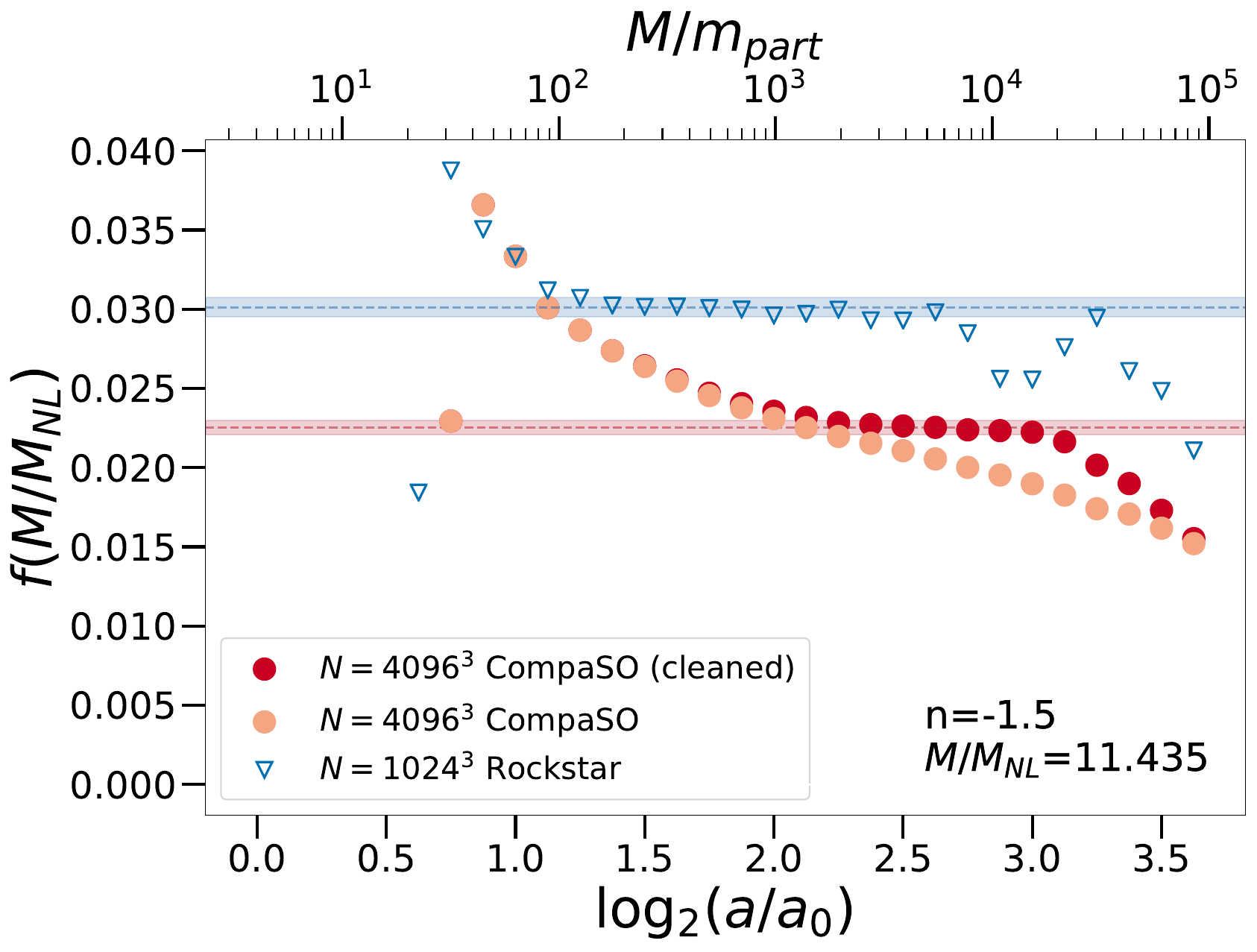}
    \end{subfigure}
    \begin{subfigure}{0.49\linewidth}%{0.24\textwidth}
    \includegraphics[width=\linewidth]{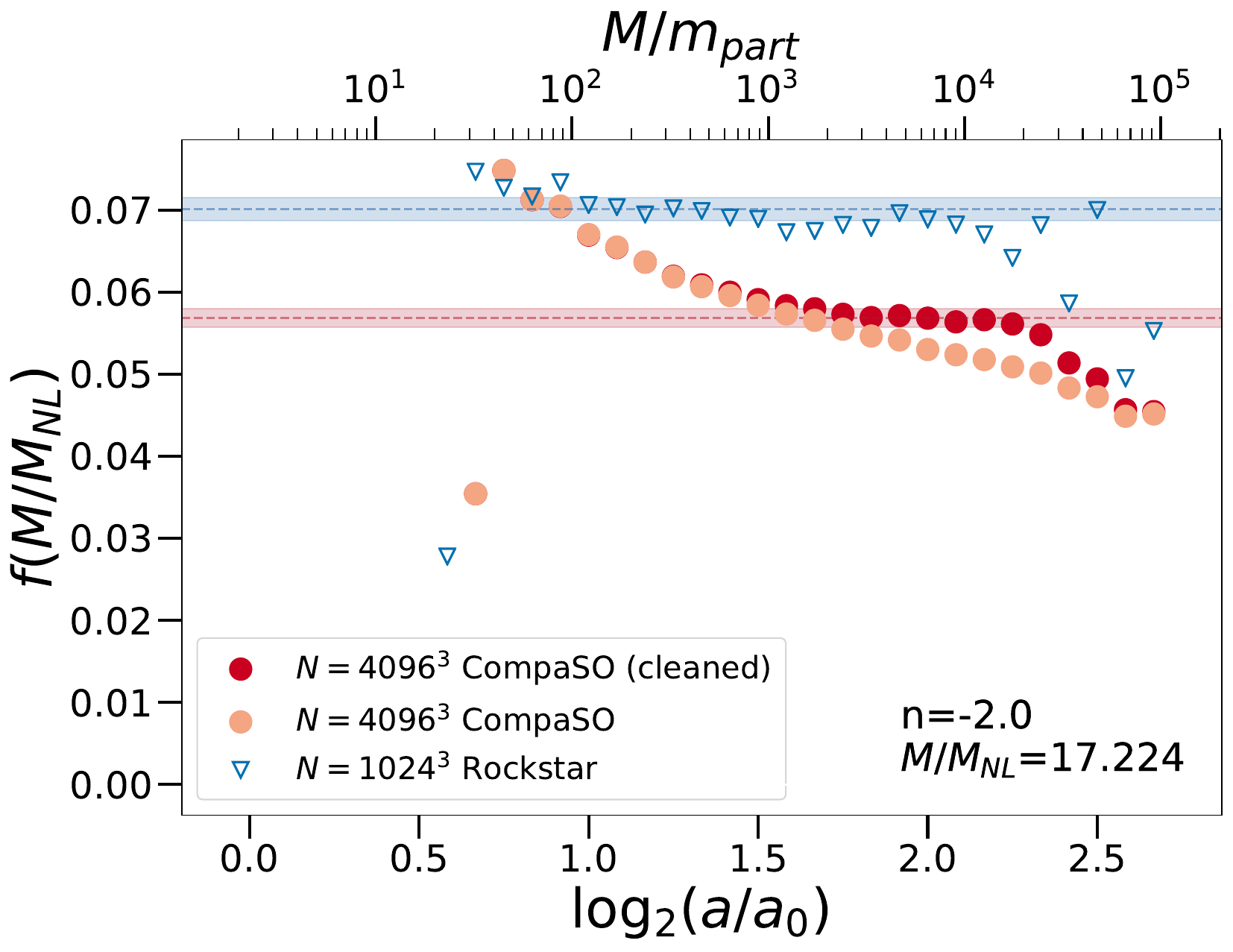}
    \end{subfigure}
\caption{Evolution of the HMF for the index $n=-1.5$ (left column) and $n=-2.0$ (right column) as a function of $\log_2(a/a_0)$, \emph{lower x-axis}, and halo particle number ($M/m_{\text{part}}$), \emph{upper x-axis}, for a set of given mass-rescaled bins $M/M_{\rm NL}$.
Blue triangles correspond to \Rockstar for a single $N=1024^3$ simulation, while circles correspond to \CompaSO for the $N=4096^3$ simulation (orange corresponds to results before merger-tree cleaning and red corresponds to results after). Horizontal dashed lines represent the converged value of the HMF, and the shaded regions indicate that within $\pm1\%$ of this value.}
\label{fig:conv_HMF}
\end{figure*}

Results for the comparison between \Rockstar and \CompaSO (before and after performing the cleaning process) are shown in \autoref{fig:conv_HMF}, for $n=-1.5$ (left panels) and $n=-2.0$ (right panels). In this figure, each plot shows the results of all halo-finders analysed at three chosen representative rescaled mass bins. We also indicate, on the upper $x$-axis, the number of particles in a halo ($M/m_{\text{part}}$), where $m_{\text{part}}$ is the particle mass of our simulations, on the upper $x$-axis. The horizontal lines in the panels indicate the estimated converged value when such convergence is attained at $1\%$ precision, using the criteria as detailed in \autoref{sec:converge_method} (with $p=0.01$). The uppermost two panels correspond to the smallest rescaled mass at which such convergence is obtained (for at least one of the finders), and the bottom panels to the largest such rescaled mass. This value of $1\%$ is chosen because it is approximately the smallest value of $p$ for which we obtain a significant range of contiguous bins satisfying our convergence criteria.

As $M_{\rm NL}$ grows as a function of time, the halos populating a given $M/M_{\rm NL}$ bin contain more and more particles as time progresses. Thus, each plot effectively shows the measured mass function as a function of increasing resolution. At the same time, as halos get bigger, their number density gets smaller. The average number of halos in each bin decreases monotonically: this number is proportional to the simulation volume in units of the characteristic volume $R_{\text{NL}}^3$, meaning that in the approximation of $M_{\text{NL}}^2n(M,a)$ constant, it's proportional to $1/M_{\text{NL}}$. Thus, we expect the effects of sparseness of sampling in finite bins to make the signal noisy at late times. We indeed can see this effect in our results, in \autoref{fig:conv_HMF} for the HMF, but further in the analysis also for the 2PCF and the pairwise velocity. Nevertheless, we see that most of our plots are clearly not dominated by such sampling noise, and we can clearly identify systematic dependences in resolution alone. We note that the different halo finders have different mass definitions, so in these figures we do not expect agreement in the value of $f(M/M_{\rm NL})$ \cite[already reported in][]{CompaSO} but we are interested instead in comparing the time/particle number range in which a convergence to a constant behaviour (i.e. self-similarity) is attained. 
  
Examining these plots, we see several clear trends depending on the halo-finder. \Rockstar catalogues show generally good convergence: looking at \autoref{fig:conv_HMF}, we see that a $1\%$ precision level is attained starting from the order of $100$ particles (regardless of the spectral index $n$), with degrading convergence at 
larger mass/later time due to sampling and smaller box size. Regarding \CompaSO halos, they show equally good convergence as \Rockstar ($1\%$ precision) beyond $\sim1000$ particles when the cleaning is performed, while 
the raw \CompaSO catalogues never meet the convergence criteria and show instead a clear monotonic dependence on the resolution. On one hand, the larger number of particles needed for convergence in \CompaSO is expected, as the kernel density scale is fixed and does not scale self-similarly (i.e., a new scale is introduced in the problem, which is expected to break self-similarity at low mass). On the other hand, the behaviour displayed by the raw \CompaSO is very similar to that observed in \citet{Leroy2021} for 
FoF-selected halos.  
Thus, the  merger-tree based cleaning (discussed in \autoref{sec:HaloFinder} above) appears to
correct very appropriately the mass of halos, by increasing by the right amount the number of larger halos at each 
given time to restore the self-similarity. This behaviour is also expected, as the described cleaning method affects, in general, low-mass halos around more massive ones, appending their particle list to the latter, and resulting in cleaned halo catalogues with a lower number of smaller halos vs. a larger number of bigger halos. As a result, this shifts the value of the HMF in each mass-bin exactly in the correct direction to preserve self-similarity, which provides clear evidence for the good performance of the procedure.  

The panels of the bottom row in \autoref{fig:conv_HMF}, which probe the most massive halos resolved, show
a clear upper cut-off in the convergence of the cleaned \CompaSO catalogues at a few times $10^4$ particles. Comparing with the plain behaviour 
seen in the same bin for the smaller \Rockstar boxes, which appear to show a down-turn of the data away of the 
converged value at a slightly earlier time, it appears that these deviations can be attributed to finite box size 
effects.  Further tests against larger \Rockstar boxes would be desirable to confirm this and exclude any evidence 
for residual resolution dependence in the cleaned \CompaSO, as well as test against self-similarity for different cleaning parameter values.

Indeed, we note that one of the more general conclusions we can draw is that the self-similarity tests on scale-free models are an excellent tool for testing resolution of halo finders. %Furthermore, while we do not claim these tests to be proof of correctness for halo's definition, self-similarity is a necessary evidence for it, and results can be used to place minimal convergence limits on halo finder algorithms.
Furthermore, we underline again that we do not claim that our tests allow us to say anything about whether one halo-finder is more physically relevant than the other. We say that self-similarity is a necessary condition for a good physical behaviour, although not a sufficient one, allowing us to place minimal convergence limits on halo finder algorithms.

\subsection{2PCF and radial pairwise velocities}\label{sec:R_CSO_stats}

We now turn to our analysis of the 2PCF, and the mean radial pairwise velocities of halo centres. We have first considered the HMF, as we would expect that any other halo statistics -- which are generically expected to depend on $M/M_{\rm NL}$ -- will be self-similar to a good approximation at a given rescaled $r/R_{\rm NL}$ only if the HMF is too. Amongst other considerations, we will examine below the extent to which this is the case quantitatively for the 2PCF and mean radial pairwise velocity. 

\begin{figure*}
    \begin{subfigure}{0.49\linewidth}%{0.24\textwidth}
    \includegraphics[width=\linewidth]{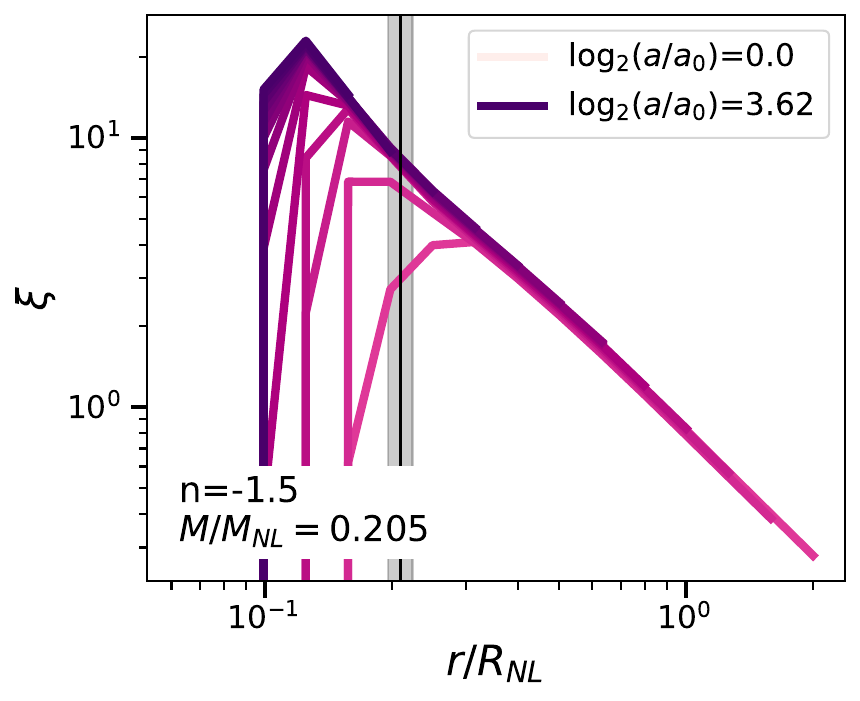}
    \end{subfigure}
    \begin{subfigure}{0.49\linewidth}%{0.24\textwidth}
    \includegraphics[width=\linewidth]{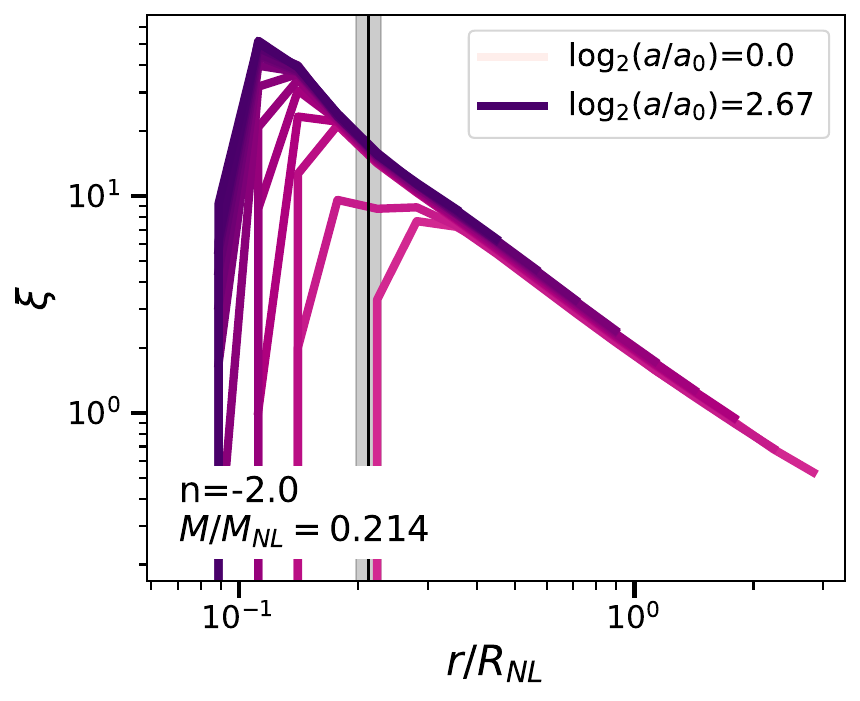}
    \end{subfigure}    
    \begin{subfigure}{0.49\linewidth}%{0.24\textwidth}
    \includegraphics[width=\linewidth]{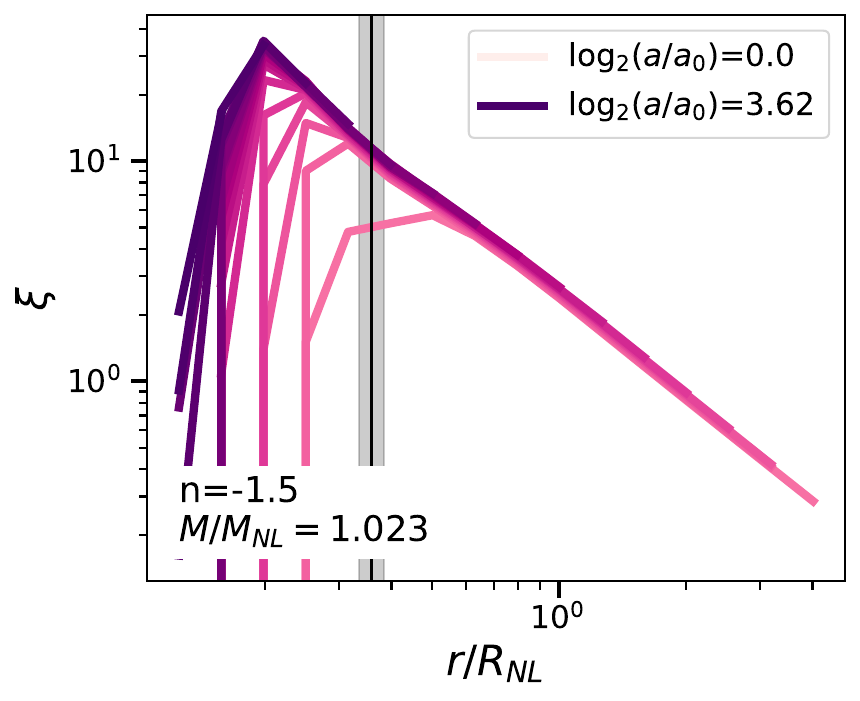}
    \end{subfigure}    
    \begin{subfigure}{0.49\linewidth}%{0.24\textwidth}
    \includegraphics[width=\linewidth]{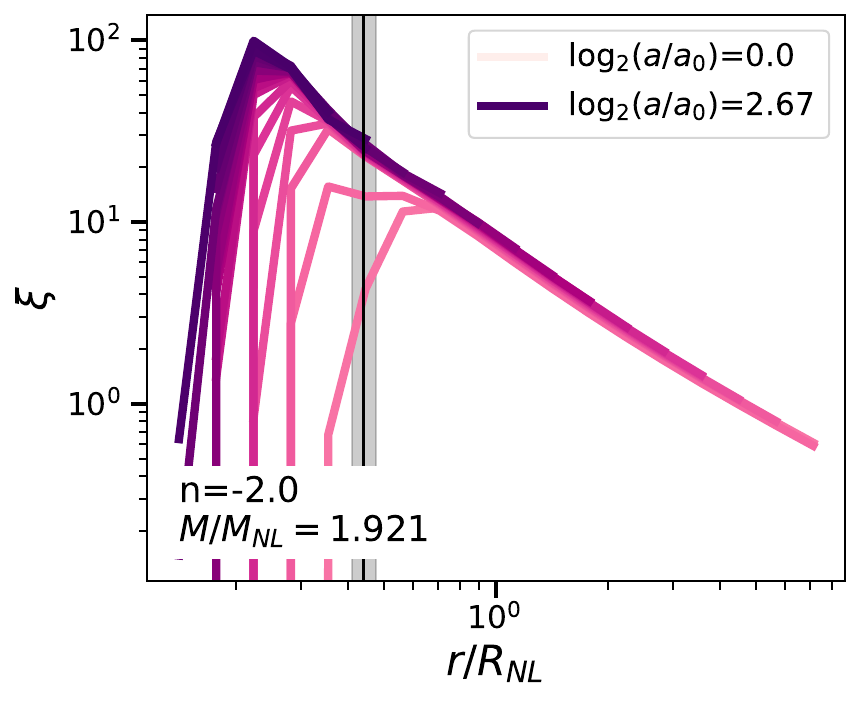}
    \end{subfigure}    
    \begin{subfigure}{0.49\linewidth}%{0.24\textwidth}
    \includegraphics[width=\linewidth]{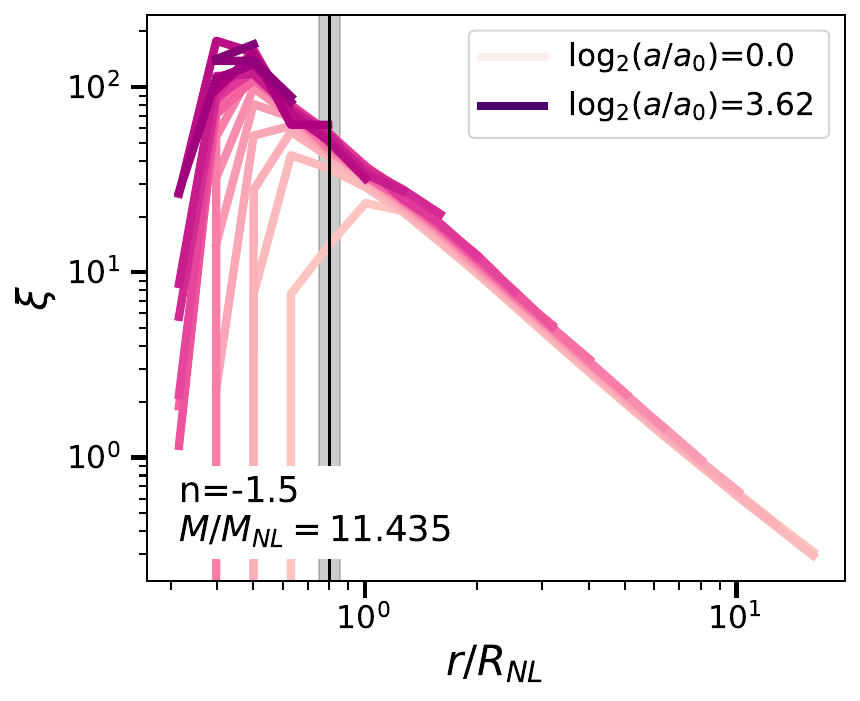}
    \end{subfigure}
    \begin{subfigure}{0.49\linewidth}%{0.24\textwidth}
    \includegraphics[width=\linewidth]{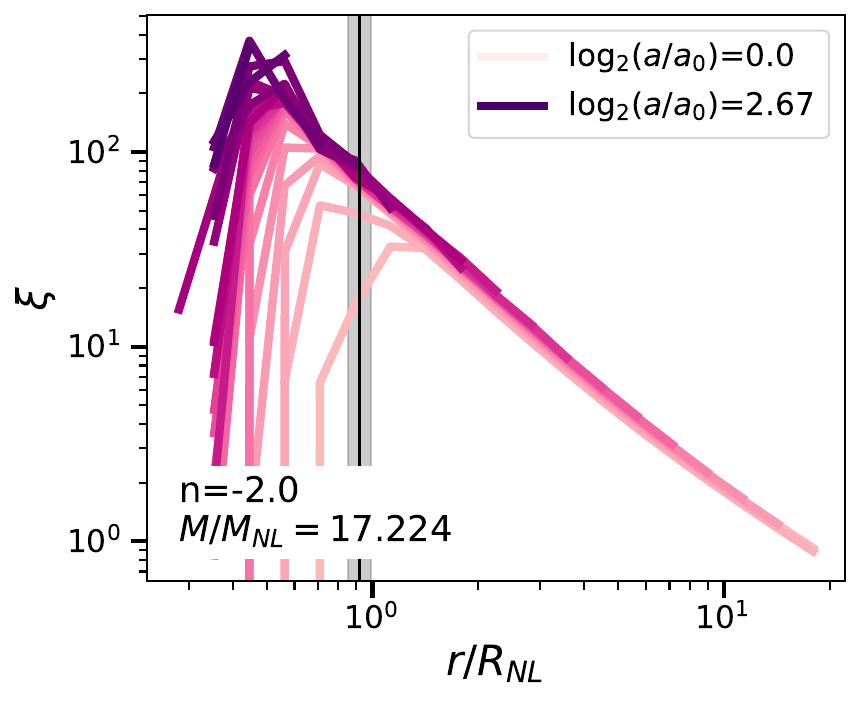}
    \end{subfigure}
\caption{2PCF as a function of rescaled separation $r/R_{\rm NL}$, for $n=-1.5$ (left column) and $n=-2.0$ (right column) simulations. The black vertical line marks twice the rescaled virial radius of the corresponding rescaled mass, while the shaded area shows the corresponding scales to the minimum and maximum mass limits on the finite bin. The statistic is computed for each bin of rescaled mass $M/M_{\rm NL}$, showing here those corresponding with \autoref{fig:conv_HMF}. The data corresponds to cleaned \CompaSO for the $N=4096^3$ simulations. For each mass bin, we show all snapshots containing halos in the bin.}
\label{fig:CF}
\end{figure*}

\begin{figure*}
    \begin{subfigure}{0.49\linewidth}%{0.24\textwidth}
    \includegraphics[width=\linewidth]{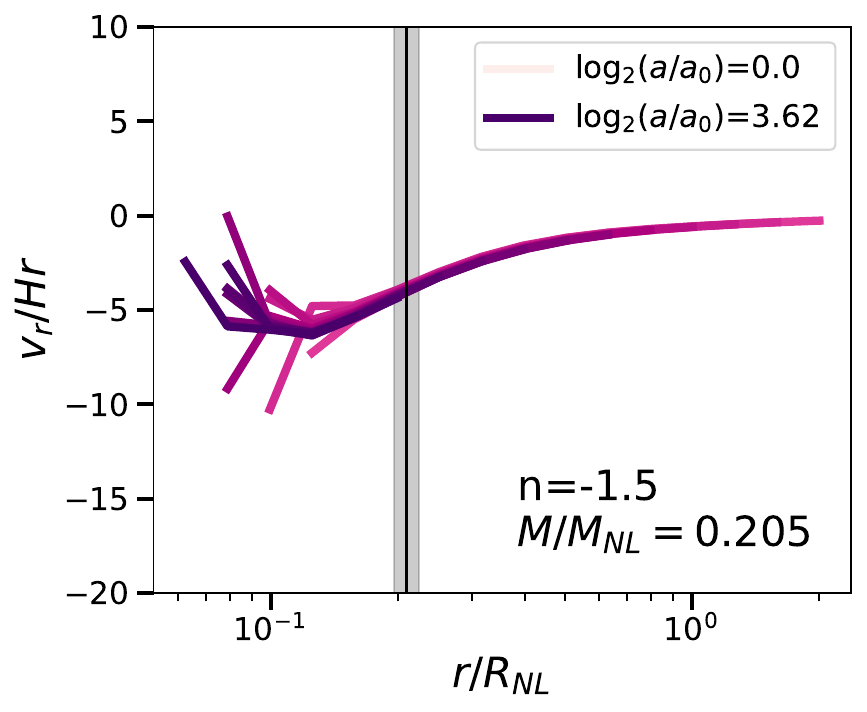}
    \end{subfigure}
    \begin{subfigure}{0.49\linewidth}%{0.24\textwidth}
    \includegraphics[width=\linewidth]{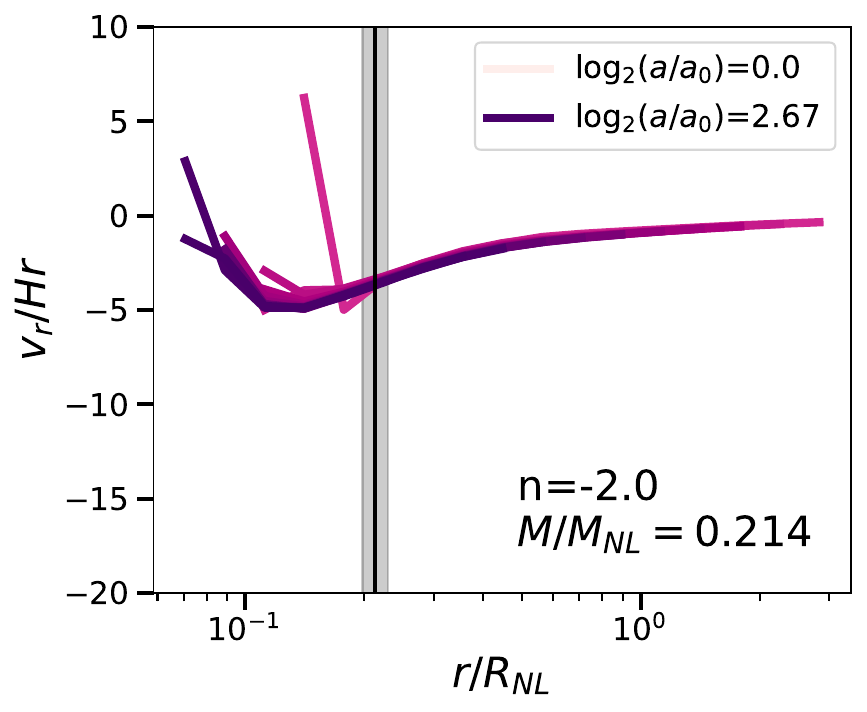}
    \end{subfigure}    
    \begin{subfigure}{0.49\linewidth}%{0.24\textwidth}
    \includegraphics[width=\linewidth]{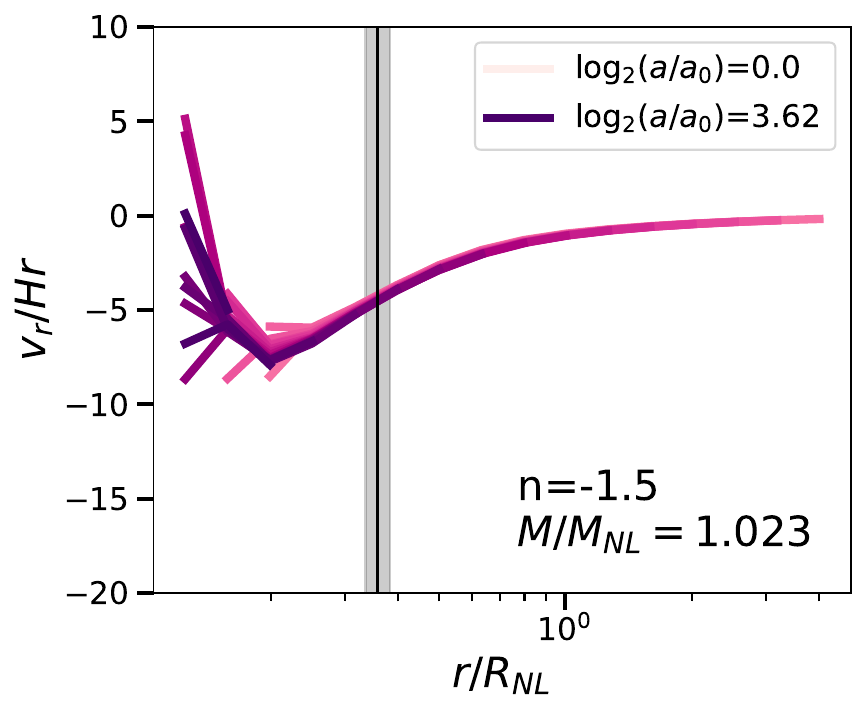}
    \end{subfigure}    
    \begin{subfigure}{0.49\linewidth}%{0.24\textwidth}
    \includegraphics[width=\linewidth]{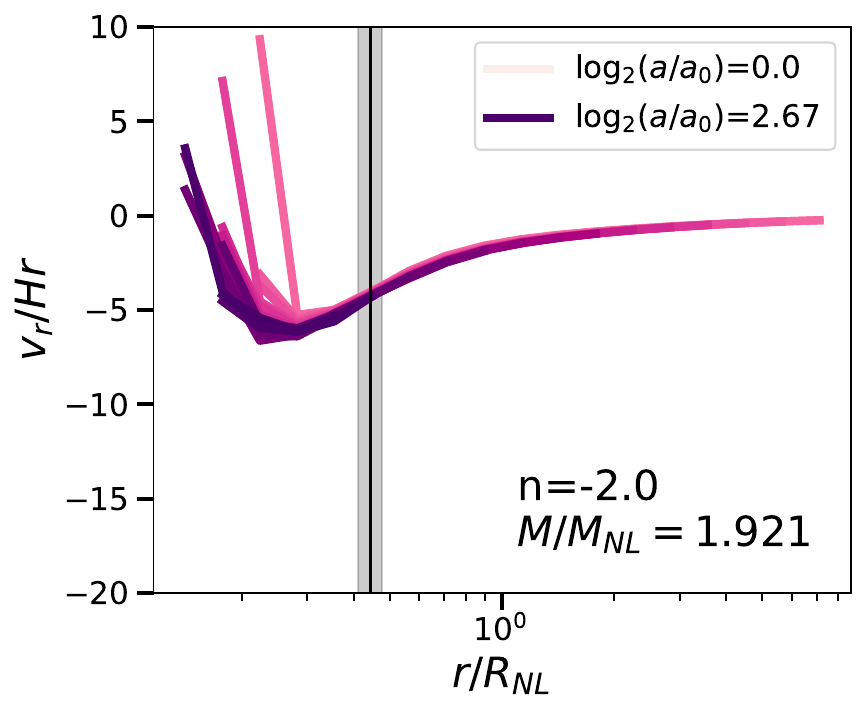}
    \end{subfigure}    
    \begin{subfigure}{0.49\linewidth}%{0.24\textwidth}
    \includegraphics[width=\linewidth]{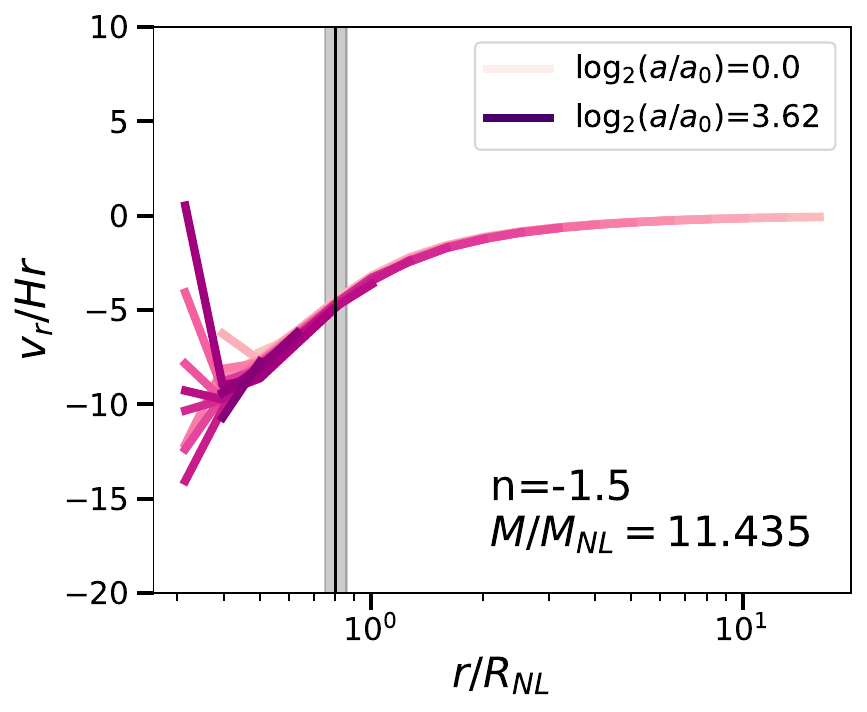}
    \end{subfigure}
    \begin{subfigure}{0.49\linewidth}%{0.24\textwidth}
    \includegraphics[width=\linewidth]{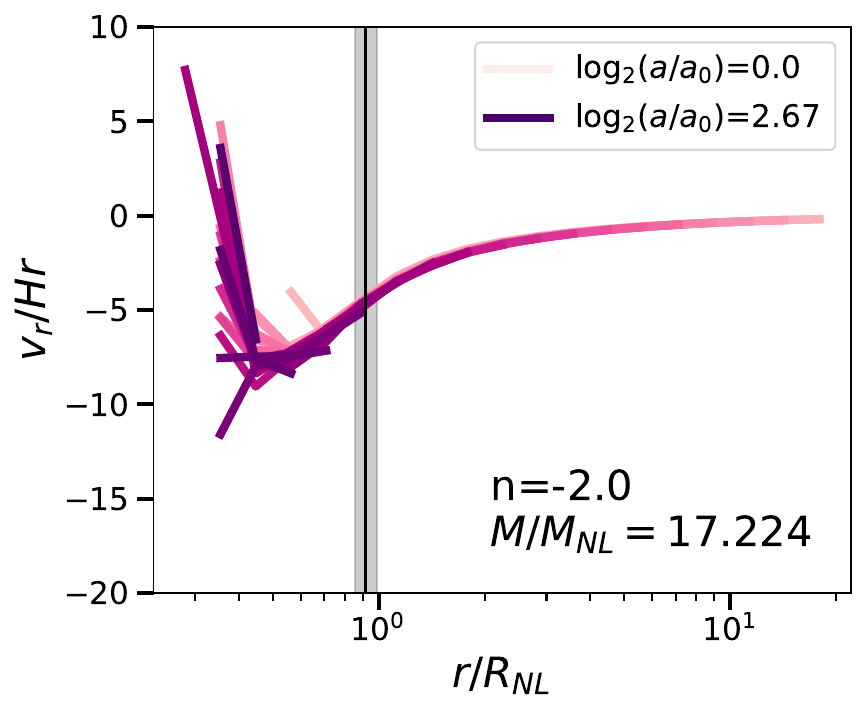}
    \end{subfigure}
\caption{Same as \autoref{fig:CF} but for the estimation of $v_r/Hr$.}
\label{fig:PW}
\end{figure*}

Looking at \autoref{fig:CF} and \ref{fig:PW}, they show, respectively, for the same three rescaled-mass bins as in \autoref{fig:conv_HMF}, the temporal evolution of the 2PCF and mean radial pairwise velocity for halo centres as a function of rescaled separation. We display results for the cleaned \CompaSO catalogues in the $N=4096^3$ simulations and the indices $n=-1.5$ and $n=-2.0$. We plot the values of the statistic as a function of $r/R_{\rm NL}$, and for all redshifts with data in the given bin of rescaled mass. 
In each plot we have marked by a black vertical line the scale $2r_{\rm vir}/R_{\rm NL}$ corresponding to twice the virial radius, $r_{\rm vir}$, of the corresponding rescaled mass.
In addition, the shaded area marks the corresponding scale to the minimum and maximum mass limits on the finite bin. Although \CompaSO halos may be separated by less than $2r_{\rm vir}$ (as they are neither spherical nor have a spatial extent directly determined by $r_{\rm vir}$) we expect 
a scale of this order to be an effective lower cut-off to the range in which a 
physical halo correlation function can be measured.

\autoref{fig:CF} and \ref{fig:PW} display qualitative behaviour similar to that in the statistics we have analysed in previous work: both statistics show clear self-similarity propagating in time from larger to smaller scales. As anticipated, the scale $2r_{\rm vir}/R_{\rm NL}$
does seem to give a good indication of the lower cut-off scale. Perhaps surprisingly, at the latest times computed for the simulations, self-similarity seems even to extend to separations as small as $r_{\rm vir}$.  
Further, the plots appear to show, again perhaps surprisingly, that the convergence of $v_r$ is slightly better than that of the 2PCF. Although the variance of $v_r$ is greater than that of the 2PCF, and thus should present a weaker convergence, we will see later on that their dependence on the HMF's convergence will play an important role on explaining this behaviour. 

\begin{figure*}
    \begin{subfigure}{0.49\linewidth}%{0.24\textwidth}
    \includegraphics[width=\linewidth]{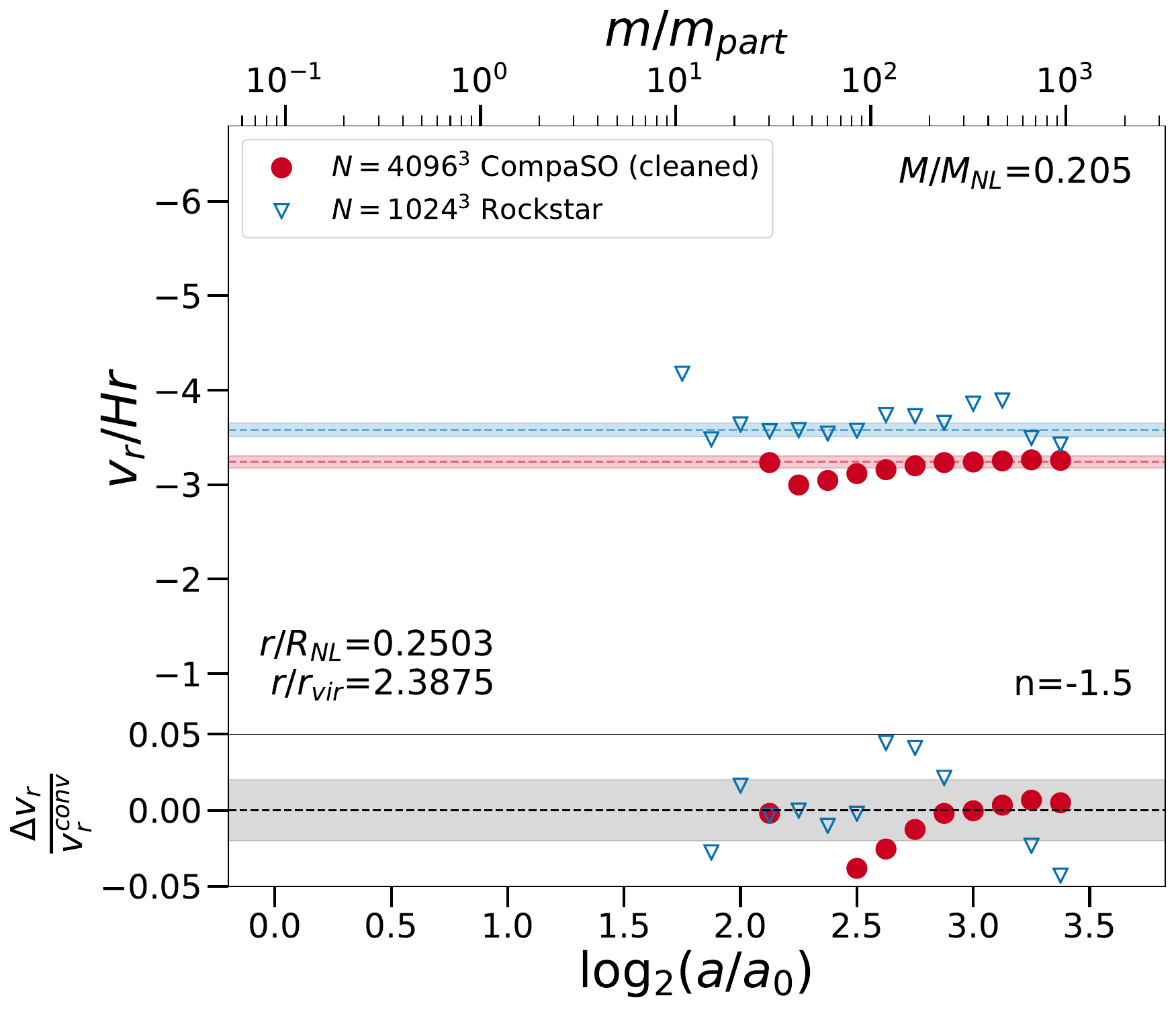}
    \end{subfigure}
    \begin{subfigure}{0.49\linewidth}%{0.24\textwidth}
    \includegraphics[width=\linewidth]{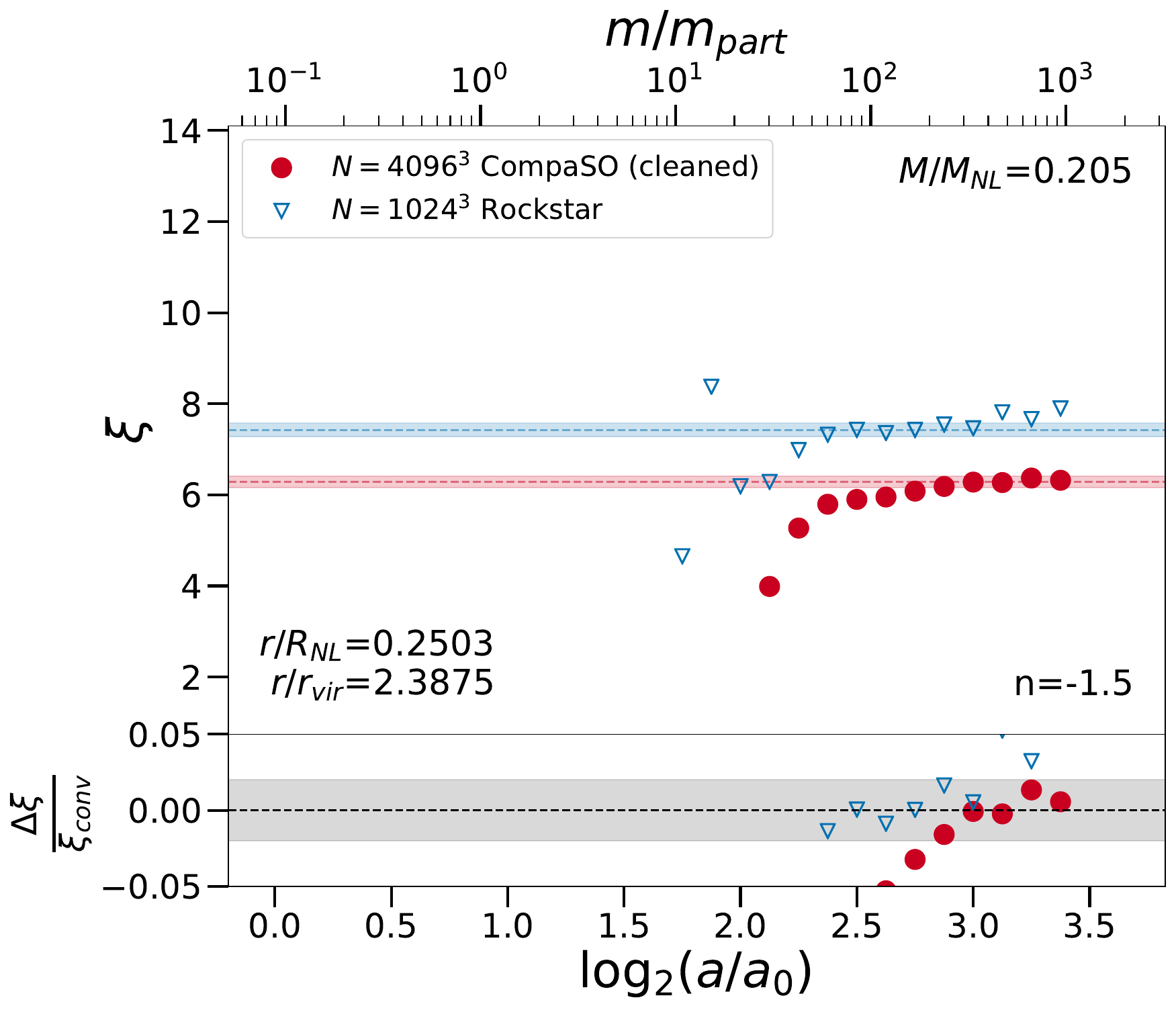}
    \end{subfigure}    
    \begin{subfigure}{0.49\linewidth}%{0.24\textwidth}
    \includegraphics[width=\linewidth]{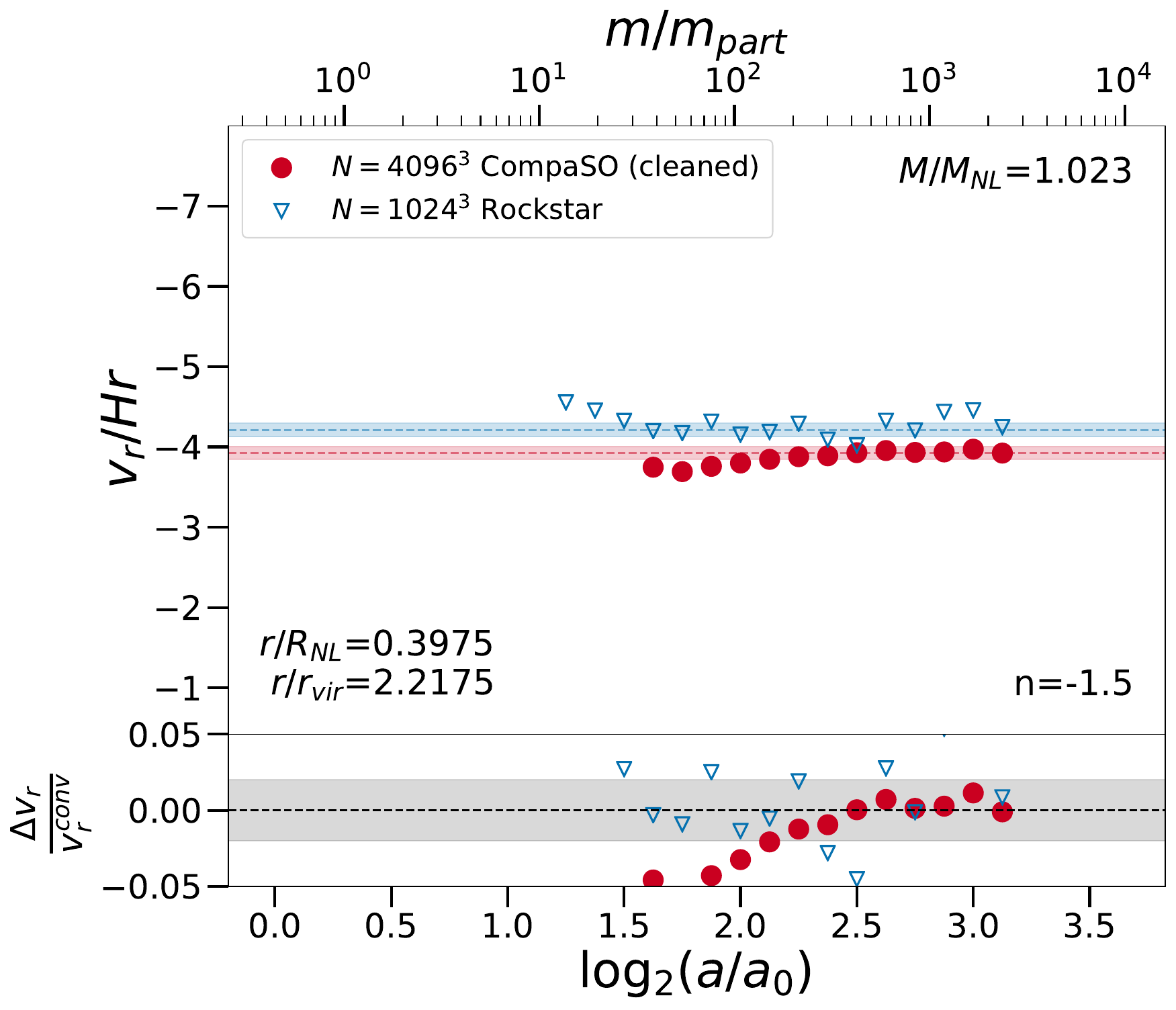}
    \end{subfigure}    
    \begin{subfigure}{0.49\linewidth}%{0.24\textwidth}
    \includegraphics[width=\linewidth]{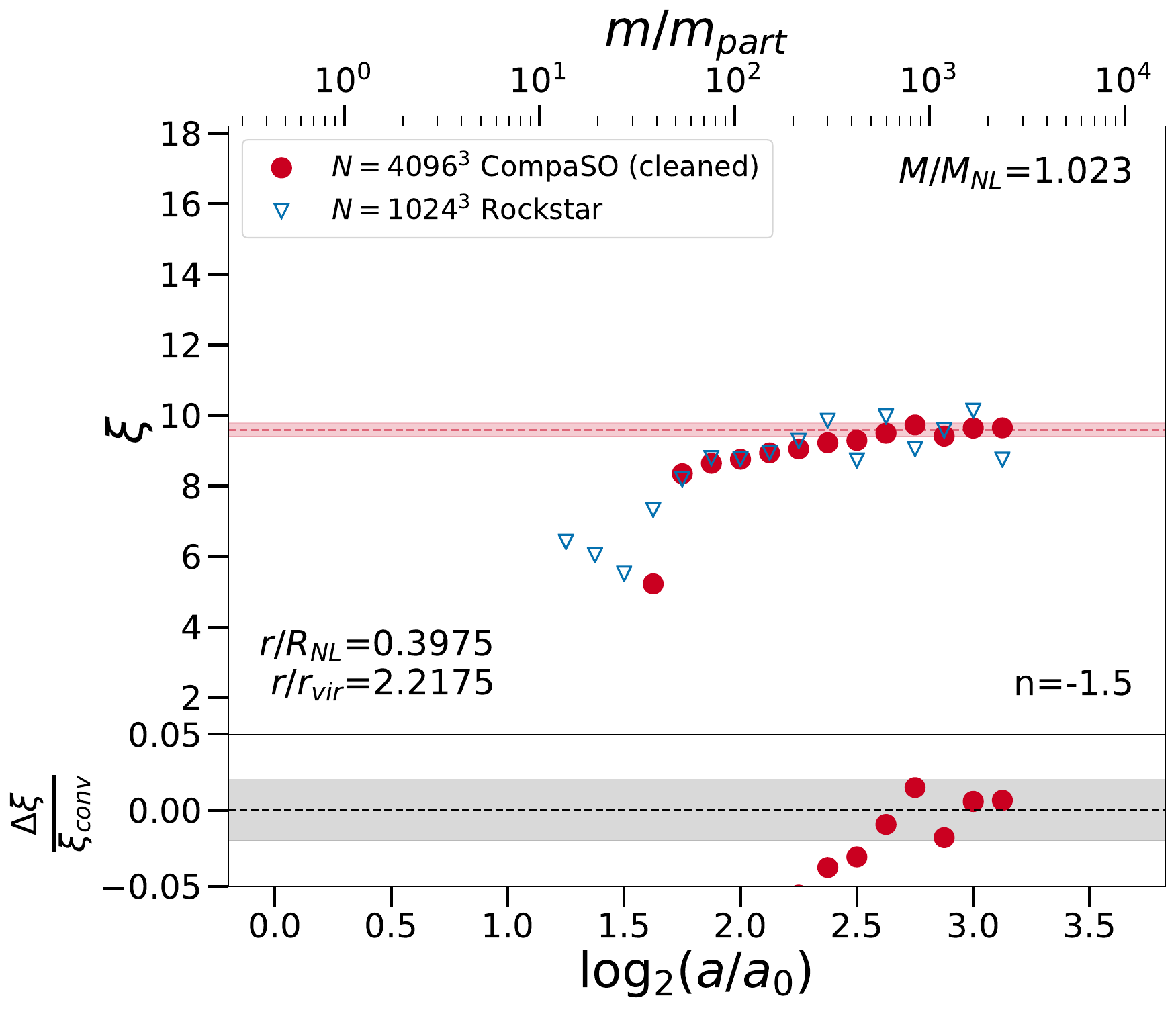}
    \end{subfigure}    
    \begin{subfigure}{0.49\linewidth}%{0.24\textwidth}
    \includegraphics[width=\linewidth]{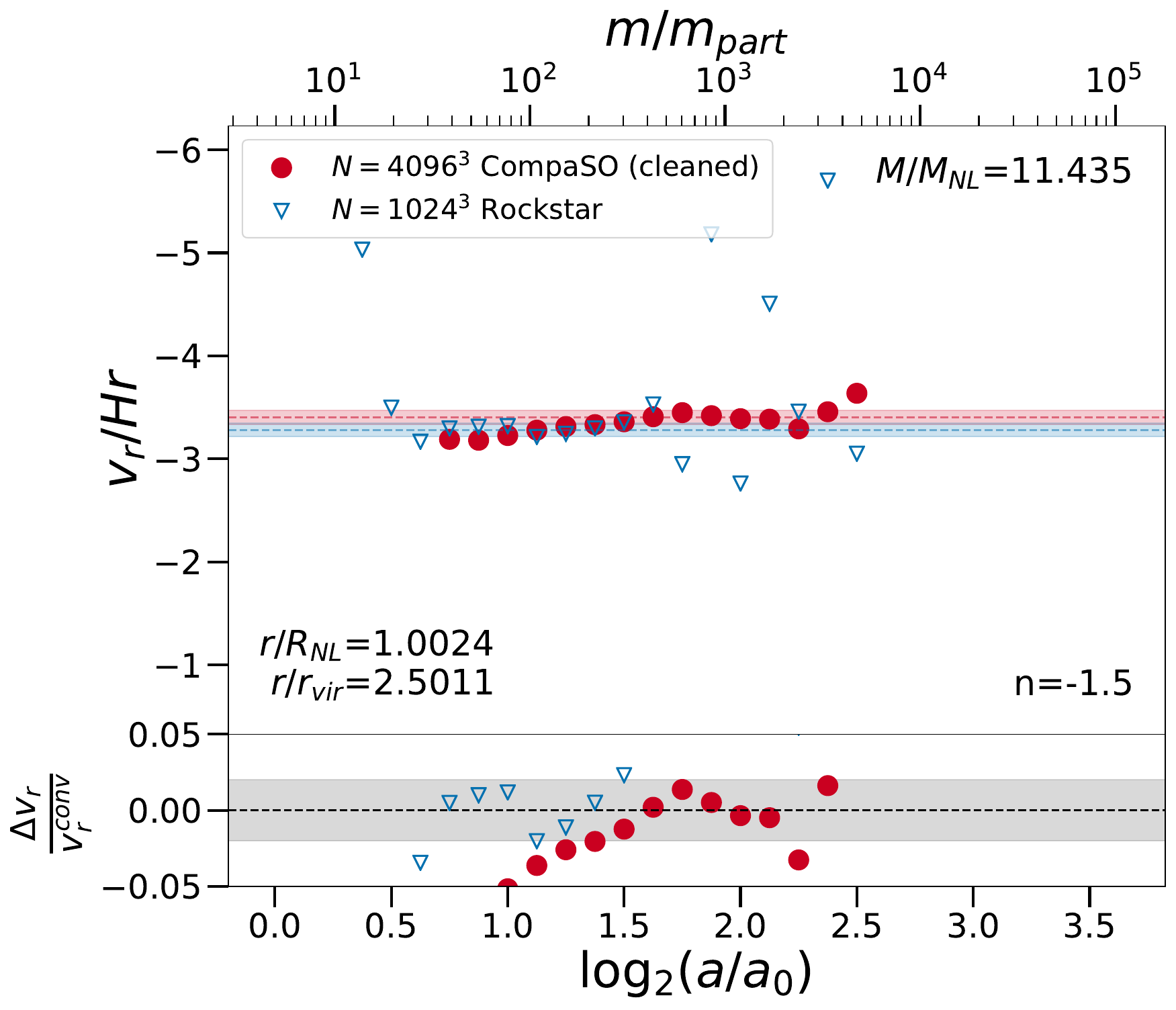}
    \end{subfigure}
    \begin{subfigure}{0.49\linewidth}%{0.24\textwidth}
    \includegraphics[width=\linewidth]{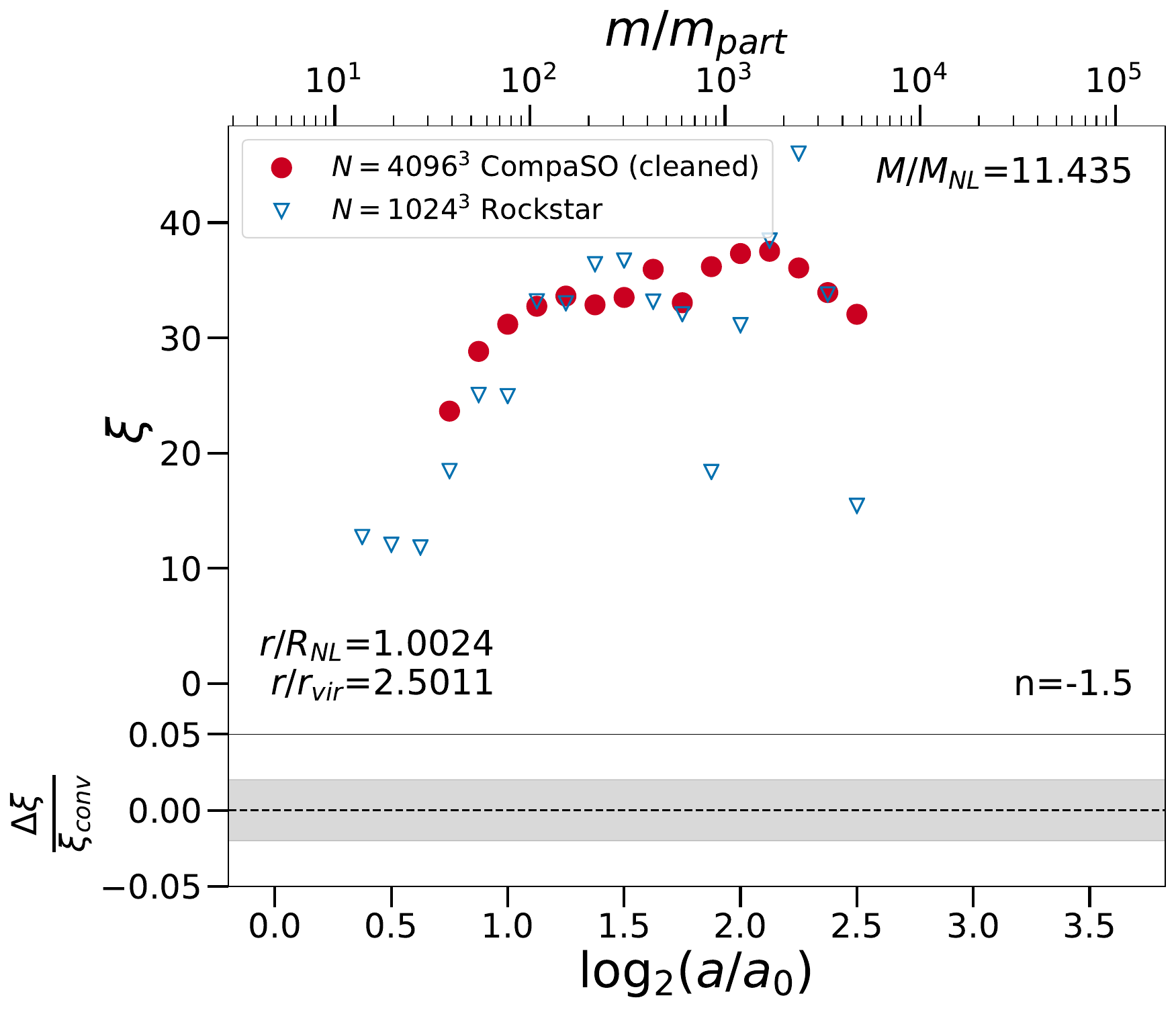}
    \end{subfigure}
\caption{Evolution, for $n=-1.5$ simulations, of $v_r/Hr$ (left column) and the 2PCF (right column) as a function of $\log_2(a/a_0)$, \emph{lower x-axis}, and number of particles in the halo, \emph{upper x-axis}. Each plot shows the scale close to $\sim2r_{\rm vir}$ for each mass-bin in \autoref{fig:HMF2}. Blue triangles correspond to \Rockstar for a single $N=1024^3$ simulation, while red circles correspond to cleaned \CompaSO for the $N=4096^3$ simulation. Horizontal red dashed lines represent the converged value of $v_{r}/Hr$ and 2PCF, and the shaded regions indicate that within $\pm2\%$ of this value. The lower panel of each plot indicates the dispersion of the direct measurement of the statistic with respect to its converged value, while the shaded region covers the imposed $\pm2\%$ precision.}
\label{fig:n15_conv}
\end{figure*}

Following the analysis detailed in \autoref{sec:converge_method}, to assess and quantify these behaviours, we take vertical slices in \autoref{fig:CF} and \ref{fig:PW}.
As $v_r/Hr$ and $\xi$ are each functions of the two rescaled variables $r/R_{\rm NL}$ and 
$M/M_{\rm NL}$, each such plot in \autoref{fig:n15_conv} thus corresponds now to a specific bin of each of these two 
variables (and self-similarity, again, to a time independent behaviour of the dimensionless statistics). Limitations of space here impose the choice of a few illustrative 
values of $r/R_{\rm NL}$ and $M/M_{\rm NL}$. 

In \autoref{fig:n15_conv} we show three
plots for each of the two statistics, for \Rockstar and cleaned \CompaSO
halo catalogues obtained in the $N=1024^3$ and $N=4096^3$ simulations of $n=-1.5$, respectively. The bins 
correspond to the same three values of $M/M_{\rm NL}$ as in \autoref{fig:conv_HMF}, \ref{fig:CF} and \ref{fig:PW}, matching the bins in which we 
obtain a satisfactory converged HMF. The statistics are converged at the 
$2\%$ level, and the converged values are indicated by horizontal lines, with the allowed precision marked by the shaded regions. 
The value of $r/R_{\rm NL}$ in each bin has been chosen to correspond 
approximately to $2r_{\rm vir}/R_{\rm NL}$, which is roughly the 
smallest scale from which we observe convergence of both statistics 
(using the same criteria). Just as in the plots for the HMF in 
\autoref{fig:conv_HMF}, we also plot 
in the upper x-axis the number of particles in the analysed halos as a function of time. We do not 
display the results for the raw \CompaSO catalogue because this data is almost exactly superimposed on that for the cleaned catalogue for $v_r/Hr$: differently to what we observed for the HMF (and for the 2PCF), the accuracy of this halo statistic, and indeed their convergence (see below) is insensitive to the associated re-assignment of particles. The value of $2\%$ precision has (like in
the corresponding HMF plots above) been chosen because it is approximately the smallest value of $p$ for which we obtain a significant range of contiguous bins satisfying our convergence criteria, corresponding to the highest precision at which we can in practice establish convergence using our data.

As anticipated from \autoref{fig:CF} and \ref{fig:PW}, the convergence of the pairwise velocity (left panels) is indeed significantly better than that of the 2PCF (right panels).
Convergence is attained (at a given precision, here $2\%$) starting from a smaller particle number (i.e. earlier in time). This difference becomes more pronounced in the largest mass bin, as clearly illustrated here in the chosen bin (bottom plots in the figure) in which the 2PCF alone fails to meet the convergence criteria. 
We believe the explanation for this comes from the very different dependencies of the
two statistics on the rescaled mass. Comparing the converged values of the two statistics in the different mass bins in \autoref{fig:n15_conv}, we see that the pairwise velocity is only very weakly dependent on the mass
compared to the 2PCF: the former varies by only $20 \%$, while the latter changes by a factor of $5$
(as the mass itself varies by a factor of $50$). Errors in the mass assignment of the halos selected 
in a given mass bin will thus feed through to give a much larger error in the 2PCF.

Examining further the lower bounds to convergence, we observe that 
the \Rockstar data converges on small scales at fewer particles per halo than the \CompaSO, while both perform equivalently at larger scales. This is clearer in the lowest mass bin, extending to the larger mass bins, albeit somewhat obscured by the relative noisiness at larger scales of the \Rockstar data (due to smaller box size). 

\subsection{Resolution limits for halo statistics in scale-free and LCDM-type simulations}
\label{sec:resolution_halos}

As we have discussed, the lower cut-offs to convergence for the halo statistics we have analysed can be stated as cut-offs on the number of particles per halo, and in the case of the correlation functions and pairwise velocity (which depend also on separation) also in terms of a cut-off on separation in units of the virial radius. Further, in the data shown above we have seen that in practice the requirement on particle number, for a given halo finder, seems not to depend significantly on the mass bin for the HMF or the pairwise velocity at a given scale, at least for the approximately fixed separations
(in units of virial radius) which we examined.

\begin{figure*}
%    \centering
    \begin{subfigure}{\linewidth}%{\textwidth}
    \includegraphics[width=\linewidth]{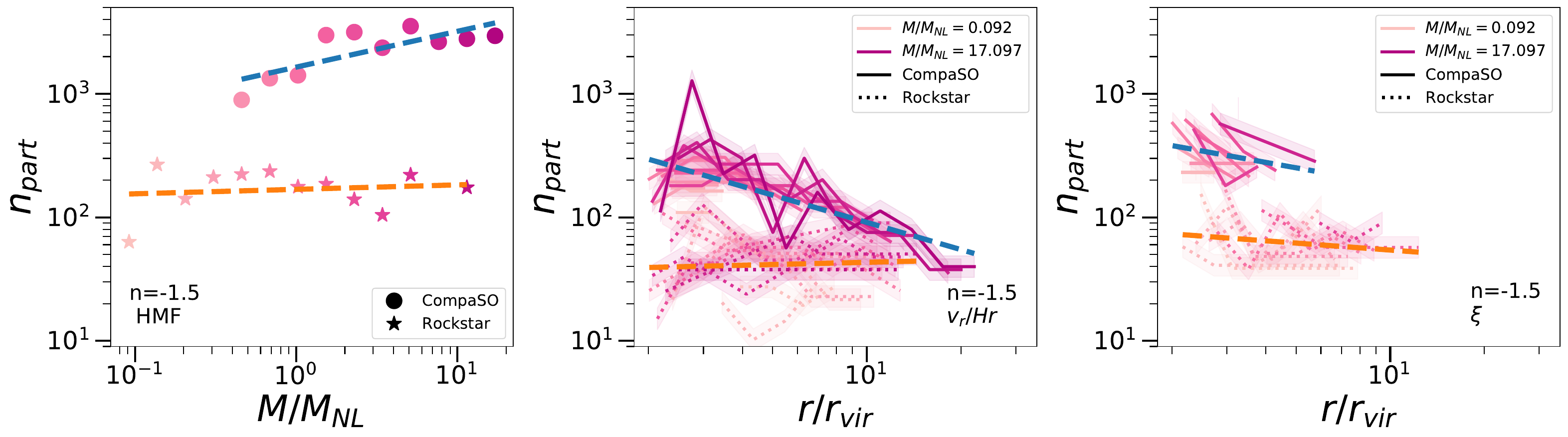}
    \end{subfigure}
    \begin{subfigure}{\linewidth}%{\textwidth}
    \includegraphics[width=\linewidth]{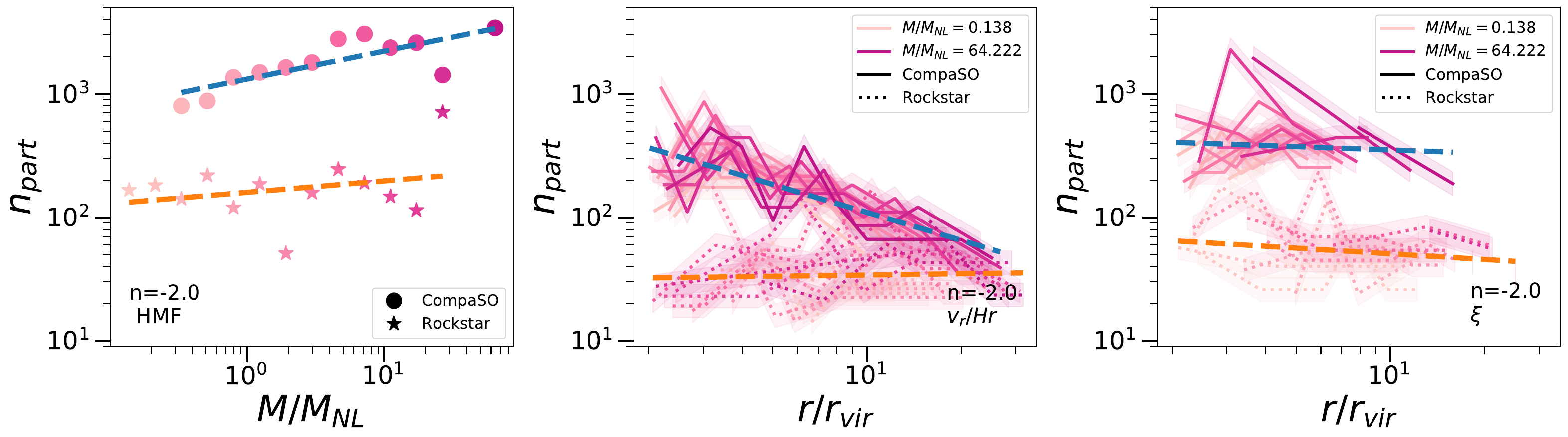}
    \end{subfigure}
    \caption{Minimum number of particles needed, at each $M/M_{\rm NL}$ bin, for convergence of: HMF (left panel), mean pairwise velocity (middle panel) and 2PCF (right panel), with the last two computed in addition at different $r/r_{\rm vir}$ bins. All results correspond to our best reported precision: $1\%$ for HMF and $2\%$ for pairwise velocity and 2PCF. Solid lines/circles correspond to cleaned \CompaSO while dashed lines/stars correspond to \Rockstar. The blue and orange dashed lines are the least squares best fit of all mass-bin data. The axes are the same in all plots to facilitate comparison.}
    \label{fig:npart_res}
\end{figure*}

\autoref{fig:npart_res} presents a more complete view of the data to test whether these behaviours are really valid in general: for $n=-1.5$ (upper panels) and $n=-2$ (lower panels). The leftmost panels show in each case the lower cut-off to convergence expressed in particles per halo for the HMF as a function of the rescaled mass, while the other two panels show, for the pairwise velocity and 2PCF respectively, the analogous quantity as a function of separation in units of $r_{\rm vir}$, and for different bins of 
rescaled mass. In each plot, the two sets of curves shown correspond to the two indicated halo finders (full line/circles to cleaned \CompaSO and dotted lines/stars to \Rockstar), and each of the curves (or points) to different mass bins $M/M_{\rm NL}$. The dashed-thick lines correspond to best (least-squares) fits of a linear dependence on $r/r_{\rm vir}$ -- $M/M_{\rm NL}$ for the HMF case -- to the data (for each of the halo-finders separately). All results correspond to our best reported precision: $1\%$ for the HMF and $2\%$ for the 2PCF and pairwise velocity. 

The plots for $v_r$ (middle panels) show that the anticipated behaviours indeed hold: the bounds 
for the different mass-bins collapse approximately onto a single line and can thus be well 
approximated as a bound on the number of particles per halo as a function of $r/r_{\rm vir}$ exclusively. On the other hand, the column (rightmost) plotting the $\xi$ data shows that, although the dependence on $M/M_{\rm NL}$ is weak for the converged values, the quality of convergence for large mass-bins (and large scales) is significantly reduced with respect to the former statistic. 

The behaviours also confirm and further quantify the trends we observed in 
\autoref{sec:R_CSO_stats}. 
In particular, we see that the number of particles per halo required for a self-similar behaviour is, for each of the two statistics, 
indeed higher for \CompaSO than \Rockstar at small scales, but this difference disappears progressively as we go to larger 
scales, where both halo finders perform similarly. We also see further 
quantified the better convergence of $v_r$ compared to $\xi$. Finally, we note that the 
actual number of particles per halo required to meet the convergence criteria for these 
two statistics are in fact considerably smaller than those required for the HMF. 
Although convergence for the latter is established at a $1\%$ and the 2-pt statistics are only converged at $2\%$ level, relaxing the precision limits for the HMF only changes very slightly the required
particle numbers.
This is explained in the same way as we explained the relative quality of the convergence 
of $v_r$ and the 2PCF: the HMF is itself a much stronger function of rescaled mass than 
the 2PCF (and a fortiori than $v_r$).   
For example, comparing the second to third rows of \autoref{fig:conv_HMF}, we see that the HMF 
changes by a factor of 10, while in  \autoref{fig:n15_conv} (as discussed above)
we see that the 2PCF varies by a factor of 3 and the mean pairwise velocity only by
$20\%$. Thus, to obtain a $2\%$ error in $v_r$ and $\xi$ we can tolerate a much larger 
error in the mass function.  It is for the same reason that $v_r$ (and $\xi$ in some range)
show no significant sensitivity to the cleaning of the
\CompaSO catalogues, as these correspond (as seen above) to
relatively small changes to the mass assigned to halos. 

Finally, we see that the two sets of plots, for $n=-1.5$ and $n=-2.0$, differ only very marginally.
Further, they are formulated in terms of mass and length units  ($M_{\rm NL}$ and $r_{\rm vir}$) 
that are also clearly defined not just in scale-free cosmologies but in any cosmology. 
Given this, it is very reasonable to take these resolution bounds to be appropriate for any 
cosmology, like LCDM, in the range in which structures are seeded by 
a linear power spectrum with a close to power law behaviour and comparable 
exponents to these. One caveat is that the scale-free models are EdS cosmologies, 
so more caution should be used when adapting the bounds at $z \approx 0$
where deviations from EdS become significant. Nevertheless, it seems unlikely 
that these effects, arising essentially from the resolution limits on
identification of halos, would have significant sensitivity to the background
cosmology.

%----------------------------------------------------------------------
\section{Conclusions}\label{sec:Conclusions}
The analysis we have reported here builds upon a set of papers \citep{Joyce2021,Garrison2021,Garrison2021c,Maleubre2022,Maleubre2023}, which have shown the usefulness of self-similarity and scale-free cosmologies in quantifying resolution of cosmological $N$-body simulations. They analysed convergence of different matter field statistics, and tested for $N$-body parameters allowing the best approximation to the continuum limit system. We have gone a step further here and expanded the analysis performed in \citet{Leroy2021}, testing resolution limits of halo-finders computed over the previously tested $N$-body simulations. Our focus
has been on the HMF, 2PCF, and radial component of the pairwise velocity, for 
halos selected using \Rockstar and \CompaSO algorithms. Compared to the aforementioned previous studies which used a single power law ($n=-2.0$) and simulations of a single size ($N=1024^3$), as in \citet{Maleubre2022,Maleubre2023} we have considered a set of both different power laws and different box sizes.  We find that self-similarity tests are indeed an excellent tool to assess the performance of different halo finders, as shown, for example, by their capacity to detect the subtle differences resulting from the cleaning of the \CompaSO catalogues. 

Our exploitation of simulations of different sizes, of several realizations, and of scale-free models with different exponents has allowed us not only to improve the results from previous work, but has also been essential for extrapolating the results to LCDM-like cosmologies.

We have been able to use our data to establish resolution limits at the $1\%$ precision level for the HMF, and at the $2\%$ level for the 2PCF and pairwise velocity in both the \Rockstar catalogues and the \CompaSO catalogues,  provided the cleaned version described in \cite{Sownak2022} is used for the HMF.
We express 
the lower limits to resolution for the HMF as a lower limit on the number of particles, which 
turns out to be roughly independent of the non-linear mass. For the pairwise velocity, which is also a
function of separation, we find that the lower bounds on the number of particles are, to a good approximation, independent of mass when plotted as a function of separation in units of the virial radius (corresponding to  
the given mass). The converged value of the 2PCF highly depends on the mass-bin analysed, and subsequently on the good categorization of halos' masses, encoded in the HMF. We thus found that convergence of the 2PCF is only attained when the HMF is converged.

Plotting the inferred lower bounds on particle numbers for each of the three statistics,
for $n=-1.5$ and $n=-2$ simulations, shows that the results have no significant dependence on $n$
and thus can be confidently adopted to LCDM-like simulations. At the $1\%$ level, \Rockstar is not able to resolve the 
HMF below $\sim100 - 200$ particles, the cleaned version of \CompaSO breaks self-similarity below $\sim1000$ particles, and its raw version never achieves this convergence at the same precision. For the 2PCF and pairwise velocities, the $2\%$ precision level is attained with significantly smaller particle numbers than the previous statistic, with the latter requiring the least. For these, the effects of cleaning \CompaSO are less significant, as the dependence on mass-bin is suppressed. At small scales, \Rockstar exhibits self-similarity starting at a smaller particle number than \CompaSO, plausibly explained by the introduction of a fixed kernel density scale in the latter, which the authors assume will certainly affect self-similarity of small objects. This difference decreases
as the separation increases and disappears at $(10-20)r_{\rm vir}$. 

We conclude by pointing out that our analysis has confirmed that self-similarity is a powerful tool to put halo algorithms to the test and compare their resolution. Once again, we underline that we cannot make any claims about which halo-finder is more physically relevant, but we can indeed put limits on their individual resolution. Finally, it would be interesting to explore, in particular, 
whether the \CompaSO algorithm can be further modified in order to improve its resolution at low halo mass, 
while maintaining its computational speed.

%----------------------------------------------------------------------

%%%%%%%%%%%%%%%%%%%%%%%%%%%%%%%%%%%%%%%%%%%%%%%%%%

\section*{Acknowledgements}

S.M. thanks Sownak Bose for guidance and technical support with the halo merger tree code for \AbacusSummit, used to \emph{clean} the \CompaSO catalogue of the scale-free simulations used in this paper. 
She also thanks the Institute for Theory and Computation (ITC) and the Flatiron Institute for hosting her in early 2022, and acknowledges the Fondation CFM pour la Recherche and the German Academic Exchange Service (DAAD) for financial support. S.M. and M.J. thank Pauline Zarrouk for useful discussions.

D.J.E. is supported by U.S.\ Department of Energy grant, now DE-SC0007881, NASA ROSES grant 12-EUCLID12-0004, and as a Simons Foundation Investigator.

This research used resources of the Oak Ridge Leadership Computing Facility at the Oak Ridge National Laboratory, which is supported by the Office of Science of the U.S. Department of Energy under Contract No. DE-AC05-00OR22725. The \AbacusSummit simulations have been supported by OLCF projects AST135 and AST145, the latter through the U.S.\ Department of Energy ALCC program.

%%%%%%%%%%%%%%%%%%%%%%%%%%%%%%%%%%%%%%%%%%%%%%%%%%
\section*{Data Availability}

Data access for the simulations part of \AbacusSummit is available through OLCF’s Constellation
portal. The persistent DOI describing the data release is
\href{https://doi.ccs.ornl.gov/ui/doi/355}{10.13139/OLCF/1811689}. Instructions for accessing the
data are given at \href{https://abacussummit.readthedocs.io/en/latest/data-access.html}{https://abacussummit.readthedocs.io/en/latest/data-access.html}.

Data corresponding to the smaller simulations as well as the derived data generated in this research will be shared on reasonable request to the corresponding author.



\bibliographystyle{mnras}
\bibliography{Bibliography} % if your bibtex file is called example.bib

%\begin{thebibliography}{99}
%\end{thebibliography}

%%%%%%%%%%%%%%%%%%%%%%%%%%%%%%%%%%%%%%%%%%%%%%%%%%

%%%%%%%%%%%%%%%%% APPENDICES %%%%%%%%%%%%%%%%%%%%%

%\appendix



%%%%%%%%%%%%%%%%%%%%%%%%%%%%%%%%%%%%%%%%%%%%%%%%%%

% Don't change these lines
\bsp	% typesetting comment
\label{lastpage}
\end{document}